\documentclass[twocolumn,pre,letterpaper,showpacs]{revtex4}

\usepackage{amssymb,amsmath}

\RequirePackage{ifthen}

\newcommand{\pd}[3][1]{\ifthenelse{#1 = 1}{\frac{\partial #2}{\partial #3}}{\frac{\partial^#1 #2}{\partial #3^#1}}}

\newcommand{\deriv}[3][1]{\ifthenelse{#1 = 1}{\frac{d #2}{d #3}}{\frac{d^#1 #2}{d #3^#1}}}

\newcommand{\fd}[3][1]{\ifthenelse{#1 = 1}{\frac{\delta #2}{\delta #3}}{\frac{\delta^#1 #2}{\delta #3^#1}}}

\usepackage{url}
\usepackage{ifpdf}
\ifpdf
  \usepackage[pdftex]{graphicx}
  \usepackage{epstopdf}   
  \usepackage[pdftex, colorlinks]{hyperref}
\else
  \usepackage[dvips]{graphicx}
  \usepackage[dvips, colorlinks, breaklinks=true]{hyperref}
\fi

\begin{document}

\title{Effective-Temperature Induced Shear Banding in the Shear-Transformation-Zone Theory of Plasticity}
\author{Anthony~Foglia}
\email{afoglia@physics.ucsb.edu}
\affiliation{Department of Physics, University of California, Santa Barbara, California 93106, USA}
\date{\today}

\begin{abstract}
  This paper examines the stability of a previously proposed version of the shear-transformation-zone (STZ) theory of plasticity where the total STZ population is determined by an effective temperature and compares it to experimental results for a metallic glass.  In particular, the addition of effective temperature dynamics to the shear transformation zone theory leads to the existence of a range of strain rates for which the strain localizes into shear bands.  Yet while the steady-state results qualitatively agree, the instability of the dynamics of the system while loading begins better describes the experimental observations.
\end{abstract}


\maketitle

\section{Introduction}
\label{sec:introduction}

Recently, Langer\cite{Langer2004} derived a finite-temperature version of shear-transformation-zone (STZ) theory wherein the equilibrium STZ density was determined by an effective temperature and qualitatively agreed with that experimental results on metallic glasses done by Lu, et al.\cite{JLGR03}   The rheology of both is characterized by viscosity that is Newtonian at low strain rates and decreases as the strain rate increases.  In the theory, the Newtonian viscosity at low strain rates can be explained by thermal creep, while the decrease in viscosity at higher strain rates results from the stress-driven dynamics of the zones.  The experiments end at high strain rates at which the material exhibits shear localization during deformation.  Towards the end of the paper, Langer hypothesized that under certain conditions, the equations could reproduce this shear banding.  This paper is an analysis of those shear-localization properties.

Stressed metallic glasses typically fail by shear localization but the underlying physical cause is under debate.  While the theories of adiabatic shear bands, in which the shear banding instability is caused by the deformation-induced increase in the thermal temperature, have been successfully applied to crystalline metals\cite{Bai1992}, experimental reports on metallic glasses are in conflict as to whether the temperature increases produced by deformation are sufficiently high enough to reach the melting point.\cite{Wright2001a,Flores1999,Yang2004}

It has been hypothesized that the configurational degrees of freedom in a non-equilibrium system, such as a metallic glass, can be described by an effective temperature that does not equal the temperature of the heat bath.
Plastic rearrangements, such as those during deformation, could increase this effective temperature up to a value near an effective melt temperature, and that could be a mechanism that induces shear localization, despite the nominal temperature of the material staying below the melting point.  This will be shown to be the case for the STZ model described in \cite{Langer2004}.

Such an effective temperature would play a similar role as the free volume in other shear transformation zone and flow-defect models, such as Spaepen\cite{Spaepen1977}---a state variable describing the disorder of a system---but need not be physically identical.  Recent simulations by Falk and Shi\cite{Falk2003, Falk2005} have suggested that fluctuations of the free volume may be be more important than its absolute value in differentiating regions of shear localization from the bulk.  The effective temperature in the STZ model is a way of parametrizing these fluctuations.

In addition to steady-state shear bands, as discussed in~\cite{Langer2004}, Huang et al.\ showed a shear band can form during the initial transients of a controlled-strain-rate experiment, even when the only possible steady-state solution is uniform.  While they examined Spaepen free-volume model, a similar examination of the STZ model with effective temperature will also show instabilities due to the initial response.  This type of instability will be shown to better explain the experimental results than a steady-state one.

This paper begins by quantifying the conditions for steady-state shear localization in the shear transformation zone model as defined by Langer in \cite{Langer2004} and then repeats the comparison to the experiments done on the metallic glass $\mathrm{{Z}r_{41.2}{T}i_{13.8}{C}u_{12.5}{N}i_{10}{B}e_{22.5}}$ (Vitreloy 1) by Lu, et al.\cite{JLGR03} with particular attention to the localization properties of the material.  First, I review the shear transformation zone model as described in \cite{Langer2004}.  In Section~\ref{sec:shear-local-linear-stability}, I examine the steady-state linear stability of the model and make some simple comparisons to the some of the metallic glass data.  The nonlinear contributions to the stability and the dynamics of the initial loading are included in Section~\ref{sec:shear-local-dynam}.  Finally, in Section~\ref{sec:conclusion}, the results are summarized and suggestions for future research are made.

\section{STZ Equations}
\label{sec:eoms}

STZ theory hypothesizes that the plastic strain of a material under stress is localized in small regions.  In these regions, called shear transformation zones, the local configuration of the material can be modeled as a two state system.  Under no stress, each configuration is equally stable, but when stress is applied, one orientation is preferred over the other.
The transition of a STZ from one state to the other, called a ``flip'', results in a plastic strain increase $\Delta \varepsilon$ of order 1.  
  This increment of plastic strain is assumed to be purely deviatoric and non-dilational.
Once a zone flips, it is incapable of flipping any further in that direction, an important distinction from previous models.

Consider a two-dimensional system to which is applied a deviatoric stress $\underline{\underline{\sigma}}$.  The deviatoric stress tensor can be described by a magnitude given by the second stress invariant $\sigma = \sqrt{\frac{1}{2} \sigma_{ij}\sigma_{ij}}$.  Denoting the density of zones that are oriented with (against) a positive stress as $n_+$ ($n_-$), the plastic strain rate can be written as
\begin{equation}
  \label{eq:1}
  \dot{\varepsilon}^{pl} = \frac{\Delta\varepsilon l^2}{\tau_0} \left( R(+s) n_- - R(-s) n_+ \right).
\end{equation}
where $\tau_0$ is the characteristic time scale for flipping, on the order of the atomic vibrational frequency $R(\pm s)$ describes the stress dependence of the rates of flipping $\mp \to \pm$, and $l^2$ is the area of an STZ, on the order of a couple of atoms.  The stress $s= \frac{\sigma}{\bar{\mu}}$, where $\bar{\mu}$ is, in general, a characteristic of the rates and will be, for the particular rates used in this paper, approximately the yield stress.

The populations of $n_{\pm}$ are governed by
\begin{equation}
  \label{eq:2}
  \tau_0 \dot{n}_{\pm} = R(\pm s) n_\mp - R(\mp s) n_\pm + \left( \Gamma + \rho \right) \left( A_{cr} - n_\pm \right).
\end{equation}
The first two terms describe the flipping between states while the last term describes the creation and annihilation of zones.  The creation and annihilation are driven by a rate determined both from the energy dissipated from plastic work, $\Gamma$,\cite{LP03} and from thermal effects, $\rho$.\cite{Langer2004, Falk2004}  The ability of the total number of STZs to change in time is what allows plastic flow at large stresses.

In the absence of stress, the STZ populations reach a steady state value of $A_{cr}$.  In \cite{Langer2004}, Langer posits that this population is described by a Boltzmann distribution $A_{cr} = \frac{n_\infty}{2} e^{-E_Z/ k_B T_{eff}}$, governed by an effective temperature $T_{eff}$, and a characteristic energy $E_Z$.  $E_Z$ is the energy required for the formation of an STZ, which is approximately the cost of a zone due to the the elastic perturbation of the surrounding material, $E_Z = k_B T_z \sim \mu l^3$, where $\mu$ is the shear modulus.

It is convenient to rewrite the STZ dynamics in terms of $\Lambda=\frac{1}{n_\infty} \left(n_+ + n_- \right)$, the total density of STZs, and $m = \frac{ n_+ - n_- }{ n_+ + n_- }$, the relative bias of STZs.  In those terms,
\begin{equation}
  \label{eq:3}
  \dot{\varepsilon}^{pl} = \frac{\varepsilon_0}{\tau_0} \Lambda \mathcal{C}(s) \left( \mathcal{T}(s) - m \right),
\end{equation}
where $\varepsilon_0 = \Delta\varepsilon l^2 n_\infty$ and
\begin{align}
  \mathcal{C}(s) &= \frac{1}{2} \left( R(+s) + R(-s) \right) \\
  \mathcal{T}(s) &= \frac{R(+s) - R(-s)}{R(+s) + R(-s)}.
\end{align}

The total STZ population $\Lambda$ evolves according to
\begin{equation}
  \label{eq:4}
  \dot{\Lambda} = \frac{1}{\tau_0} \left( \Gamma + \rho \right) \left( e^{-1/\chi} - \Lambda \right),
\end{equation}
where $\chi = \frac{T_{eff}}{T_Z}$.  It will be shown in Section~\ref{sec:chi-dynamics}, when the effective temperature dynamics are detailed, that the timescale associated with the equilibration of the STZ density to $e^{-1/\chi}$ is much faster than that associated with the effective temperature changes.  Ergo for the present purposes we shall assume $\Lambda = e^{-1/\chi}$ at all times.  In this limit, the $\dot{m}$ equation simplifies to
\begin{equation}
  \label{eq:5}
  \dot{m} = \frac{1}{\tau_0} \left[ 2 \mathcal{C}(s) \left( \mathcal{T}(s) - m \right) - m \left( \Gamma + \rho \right) \right].
\end{equation}

Following the logic of \cite{Pechenik2005}, and done in detail in \cite{Langer2004}, the dissipation $\Gamma$ can be determined from energy balance arguments.  Balancing the rate of plastic work with the rate of change of the recoverable energy stored in the STZs $\psi$ and the dissipation yields
\begin{equation}
  \label{eq:6}
  2 \sigma \dot{\varepsilon}^{pl} = \deriv{\psi}{t} + \bar{\mu} \frac{\varepsilon_0}{\tau_0} \Lambda \Gamma.
\end{equation}
Using the fact that the dissipated energy should always be positive shows $\Gamma$ to have the form
\begin{multline}
  \label{eq:7}
  \Gamma(s,m)= \\
  \frac{ 2 \mathcal{C}(s) \left( \mathcal{T}(s) - m \right) \left( s - \mathcal{T}^{-1}(m) \right) + m \rho \mathcal{T}^{-1}(m) }{ 1 - m \mathcal{T}^{-1}(m) }.
\end{multline}
Since $R(-s)$ is a monotonically increasing function of $s$, $\mathcal{T}(s)$ will be also, therefore $\mathcal{T}^{-1}(m)$ is both well-defined and monotonically increasing.  This creates a dynamical upper bound on $m$ where $m \mathcal{T}^{-1}(m) = 1$ and the dissipation becomes infinitely large.  Combining Eqs.\eqref{eq:5} and \eqref{eq:7} yields
\begin{equation}
  \label{eq:8}
  \dot{m} = \frac{1}{\tau_0} \frac{ 2 \mathcal{C}(s) \left( \mathcal{T}(s) - m \right) \left( 1- m s \right) - m \rho }{ 1 - m \mathcal{T}^{-1}(m) }.
\end{equation}

For a system under controlled stress, Eq.~\eqref{eq:8} is solvable, yielding a steady-state value of $m$
\begin{multline}
  \label{eq:9}
  m_{ss}(s) = \frac{1}{2 s} \left[ \left( 1 + s \mathcal{T}(s) + \frac{ \rho }{ 2 \mathcal{C}(s) } \right) \vphantom{- \sqrt{ \left( 1 + s \mathcal{T}(s) + \frac{ \rho }{ 2 \mathcal{C}(s) } \right)^2 - 4 s \mathcal{T}(s) }}\right.\\
  \left. \vphantom{ \left( 1 + s \mathcal{T}(s) + \frac{ \rho }{ 2 \mathcal{C}(s) } \right)} - \sqrt{ \left( 1 + s \mathcal{T}(s) + \frac{ \rho }{ 2 \mathcal{C}(s) } \right)^2 - 4 s \mathcal{T}(s) } \right].
\end{multline}

In the limit $\rho \to 0$, where thermal effects are absent, Eq.~\eqref{eq:9} simplifies to the same results as the athermal STZ model: a family of jammed solutions where $m = \mathcal{T}(s)$ and $\dot{\varepsilon}^{pl} = 0$ for $|s| < s_y$ and a family of STZ flowing solutions where $m = \frac{1}{s}$ and $\dot{\varepsilon}^{pl} \ne 0$ for $|s| > s_y$ where $s_y$ is defined by $s_y \mathcal{T}(s_y) = 1$.  The thermal effects smooth this transition by causing slow plastic creep for $s<s_y$.

\subsection{Rate factors}
\label{sec:rate-factors}

To progress further, we must get more precise about the forms of the rate factors $\rho(T)$ and $R(s)$.  For the thermal frequency, $\rho(T)$ should have the form
\begin{equation}
  \label{eq:10}
  \rho(T) = e^{-\alpha(T)}.
\end{equation}

Since at the temperatures of interest, the thermal noise frequencies are considerably lower than the atomic vibrational frequency $\tau_0^{-1}$, $\alpha(T)$ will be greater than 1.  Determining the actual form of $\alpha(T)$ from a theory is beyond the scope of this paper, but to compare to experiments I shall follow the lead of \cite{Langer2004} and use a Cohen-Grest formula\cite{MHC79} as a phenomenological fit for $\alpha(T)$,
\begin{equation}
  \label{eq:11}
  \alpha(T) = \frac{ T_R }{T - T_0 + \sqrt{ \left( T - T_0 \right)^2 + T_1 T } }.
\end{equation}

For the rate factors $R(s)$, at 0 stress, any flipping should be the result of thermal kicking, and assuming the energy barrier for an STZ flip is the same as that in creation and annihilation, $R(0) = \rho(T)$.  Then the simplest way to add a well-behaved stress dependence is to claim that the activation energies in the rates depend exponentially on $s$, ensuring they are always positive.  This gives $R(s)$ the general form
\begin{equation}
  \label{eq:12}
  R(s) = \exp\!\left( - \alpha(T) \exp\!\left( - \gamma s \right) \right).
\end{equation}
For simplicity, I will make the same assumption implied in \cite{Langer2004} and \cite{Pechenik2005} and use $\gamma = 1$.  Consequently, $\mathcal{C}$ and $\mathcal{T}$ become
\begin{align}
  \label{eq:13}
  \mathcal{C}(s) &= \exp(-\alpha \cosh(s)) \cosh( \alpha \sinh(s)) \\
  \label{eq:14}
  \mathcal{T}(s) &= \tanh( \alpha \sinh(s)).
\end{align}
Remember that the STZ yield stress $s_y$ is defined to satisfy the condition that $s_y \mathcal{T}(s_y)=1$.  For $\alpha(T) \gtrsim 2$ (which for the metallic glasses will turn out to be $T< 1000$ K), $s_y \sim \bar{\mu}$ making $\bar{\mu}$ the yield stress.

  This expression for the transition rate will be adequate for the present purposes, but some small issues may arise.  One possible problem is that these rate factors lead maximum strain rate; as $s \to \infty$, $\mathcal{C}(s) \to \frac{1}{2}$ and $\mathcal{T} \to 1$, limiting the strain rate to at most $\frac{\varepsilon_0}{2 \tau_0} \exp\!\left(-\frac{1}{\chi_\infty}\right)$, where $\chi_\infty$ is the maximum value of the effective temperature and will be described in the next section.

Also the average rate factor $\mathcal{C}(s)$ is itself a sharply increasing function of stress.  It is low for small stresses then rapidly increases to $\frac{1}{2}$ when the stress is roughly $\ln(\alpha)$ 
.  As the temperature decreases, not only does $\mathcal{C}(0)$ decrease, the transition to $\mathcal{C} \approx \frac{1}{2}$ becomes sharper and occurs at a higher stress.  Eventually, in the limit $T \to 0$, the total rates will be zero for all finite stress, and such a material would exhibit no plasticity.

\subsection{Effective Temperature Dynamics}
\label{sec:chi-dynamics}
As stated earlier, the total STZ population is described by an effective temperature $\chi$.  Because $\chi$ describes the configuration of the system, it increases from plastic work, and decreases as the system over time anneals.

  This effective temperature is not simply the free volume.  In particular, some molecular dynamics simulations hint that the free volume may decrease in areas undergoing plastic strain\cite{Falk2005}.  Such local free volume decreases can still correspond to an increase of the effective temperature.

The effective temperature dynamics are determined by
\begin{equation}
  \label{eq:15}
  \dot{\chi} = \frac{\varepsilon_0}{\tau_0 c_0} \left[ \Lambda \Gamma(s,m) \left( \chi_\infty - \chi \right) + \kappa \rho e^{-\beta/\chi} \left( \frac{T}{T_Z} - \chi \right) \right],
\end{equation}
where $c_0$ is a dimensionless measure of the heat capacity in a volume the size of an STZ\@.  The first term, the heating, is proportional to the total dissipation rate from plastic deformation, which drives the effective temperature to a limiting value of $\chi_\infty = T_\infty/T_Z$.  While Langer assumed $T_\infty = T_g$, we will relax that assumption, simply claim that $T_\infty$ is on the order of $T_g$.

The second term describes the cooling, which depends on the thermal frequency $\rho$.  $\kappa$ is a free parameter that will give us some control of the interesting stress scales, as will be explained shortly.  The factor $\exp(-\beta/\chi)$ is the population of sites at which the effective temperature configurations can relax, $\beta$ being the typical energy of such a fluctuation.  These need not be STZs, but simply any high energy configuration, bistable or not.  The distinction is small, and we expect $\beta \sim 1$, but it is necessary because steady-state localization only can occur when $\beta \ne 1$.

  The $\Lambda$ dynamics can be ignored because the STZ density equilibrates with the effective temperature much faster than $\chi$ equilibrates to its final value.  The typical timescale for $\Lambda$ is $\tau_0 \Gamma^{-1}$ while for $\chi$ the timescale is $\tau_0 \left( \varepsilon_0  e^{-1/\chi} \Gamma \right)^{-1}$.  If we assume that when $T_{eff} = T_\infty$ the entire material is fully covered with STZs, then $\varepsilon_0 \sim n_\infty \sim e^{1/\chi_\infty}$, and the ratio of these two timescales is $\exp\!\left(-\frac{T_z}{T_\infty} \left( \frac{T_\infty - T}{T} \right) \right)$.  For the metallic glasses, $T_Z \sim$ 25,000 K, $T_\infty \sim$ 900 K, and the temperatures of interest in are all below 700 K, giving a ratio on the order of $10^{-4}$.  The STZ dynamics are at a much faster timescale than the effective temperature and we are justified in enforcing $\Lambda = \exp(-1/\chi)$.

The solutions to $\dot{\chi} = 0$ vs.\ stress for $\beta = 0.6$ for three different temperatures are shown in the top graph of Fig.~\ref{fig:stability_graph}.  For certain ranges of stresses, there are three different effective temperatures are steady-states, which means a uniform system can have three possible strain rates.  I expect the lowest and highest strain rates to be stable, but the middle rate unstable.  A system feeling a stress in this unstable range, and at a strain rate between these highest and the lowest strain rates, can be expected to break up into regions of fast and slow straining.  A quantitative examination is the aim of the next section.

\section{Shear Localization: Linear Stability}
\label{sec:shear-local-linear-stability}

\subsection{Theoretical Results}
\label{sec:linear-stability}

As evidenced by Figs.~10 and 11 in \cite{Langer2004}, the STZ equations Eqs.~\eqref{eq:8} and \eqref{eq:15} can, under certain conditions, have multiple steady-state solutions corresponding to differing effective temperatures and strain rates at same stress.  I now examine the model to find these necessary conditions on the parameters and state variables of a strip of material under simple controlled-shear-stress conditions such that it is unstable.  Denoting the flow direction as $x$, by symmetry the state variables $s$, $m$, and $\chi$ will be functions of $y$ only.
  We shall assume inertial effects are negligible, which will allow us to assume the stress throughout the strip to be uniform and equal to the applied stress.  Likewise, since the $m$ steady state depends solely on the stress, it too will be uniform.  Therefore any shear banding will be caused by inhomogeneities in the effective temperature.  Consequently we shall study the stability of perturbations of a uniform steady state ($s_0$, $m_0$, and $\chi_0$) of only the effective temperature.  With diffusion effects included into Eq.~\eqref{eq:15} the effective temperature obeys
\begin{multline}
  \label{eq:16}
  \dot{\chi} = \frac{\varepsilon_0}{\tau_0 c_0} \left\{ e^{-1/\chi} \Gamma(s,m) \left( \chi_\infty - \chi \right) \vphantom{ + \kappa \rho e^{-\beta/\chi} \left( \frac{T}{T_Z} - \chi \right) + D_\chi \pd[2]{\chi}{y}} \right. \\
  \left. \vphantom{ e^{-1/\chi} \Gamma(s,m) \left( \chi_\infty - \chi \right)} + \kappa \rho e^{-\beta/\chi} \left( \frac{T}{T_Z} - \chi \right) + D_\chi \pd[2]{\chi}{y} \right\}.
\end{multline}
Note that we only are considering diffusion perpendicular to the flow.  For particles in a simple shear flow, the diffusion along the flow is enhanced by Taylor dispersion effects which are beyond the scope of this paper.

Consider a perturbation the effective temperature of form $\chi = \chi_0 + \chi_1 e^{iky}e^{\omega(k) t}$, of a system under controlled uniform stress $s_0$, and with the uniform, steady-state state variables $m_0$ and $\chi_0$.  To zero-th order, all of the time derivatives [Eqs.~\eqref{eq:8} and \eqref{eq:16}] are equal to 0.  To first order, the perturbation is governed by
\begin{multline}
  \label{eq:17}
  \omega(k) \chi_1 = \frac{\varepsilon_0}{\tau_0 c_0} \left\{ e^{-\frac{1}{\chi_0}} \Gamma_0 \left( \frac{1}{\chi_0^2} \left( \chi_\infty -\chi_0 \right) - 1 \right) \right.\\
    \left. + \kappa \rho e^{-\frac{\beta}{\chi_0}} \left( \frac{\beta}{\chi_0^2} \left( \frac{T}{T_Z} -\chi_0 \right) - 1 \right) - D_\chi k^2 \right\} \chi_1,
\end{multline}
where $\Gamma_0 = \Gamma(s_0,m_0)$.  Since $\deriv{\chi_0}{t} = 0$, $\Gamma_0 = \kappa \rho e^{-\left(\beta - 1\right)/\chi_0} \left( \frac{ \chi_0 - \frac{T}{T_z} }{\chi_\infty - \chi_0} \right)$.  Using that to remove any explicit dependence on the stress in the stability exponent $\omega(k)$ yields
\begin{multline}
  \label{eq:18}
  \omega(k) = \frac{\varepsilon_0}{\tau_0 c_0} \left\{ \kappa \rho e^{-\beta/\chi_0} \vphantom{ \left[ \frac{1-\beta}{\chi_0^2} \left( \chi_0 - \frac{T}{T_Z} \right) - \frac{ \chi_\infty - \frac{T}{T_Z} }{ \chi_\infty - \chi_0 } \right] - D_\chi k^2} \right. \\
  \left. \times \left[ \frac{1-\beta}{\chi_0^2} \left( \chi_0 - \frac{T}{T_Z} \right) - \frac{ \chi_\infty - \frac{T}{T_Z} }{ \chi_\infty - \chi_0 } \right] - D_\chi k^2 \right\}.
\end{multline}

The most unstable mode is a uniform effective temperature change ($k=0$).  Generally, the effective temperature will be between $T_\infty$ and $T$, which, in the case where $T_\infty > T$, means that $\beta > 1$ is required for instability.  More precisely, there will only be a region of instability for material parameters satisfying
\begin{equation}
  \label{eq:19}
  1 - \beta > \frac{ 4 \frac{T}{T_Z} \chi_\infty }{ \chi_\infty - \frac{T}{T_Z} }.
\end{equation}

If this condition is satisfied, there will be a region of instability for effective temperatures between $\chi_\pm$ given by
\begin{multline}
  \label{eq:20}
  \chi_\pm = \frac{1}{1+\frac{2}{1-\beta} \Delta\chi} \\
  \times \left[ \bar{\chi} \pm \sqrt{\bar{\chi}^2 - \frac{T}{T_Z}\chi_\infty \left( 1 + \frac{2}{1-\beta} \Delta\chi \right) } \right],
\end{multline}
where $\bar{\chi} = \frac{1}{2} \left( \chi_\infty + \frac{T}{T_Z} \right)$ and $\Delta\chi = \frac{1}{2} \left( \chi_\infty - \frac{T}{T_Z} \right)$.

Interestingly, the bistability conditions are independent of $\kappa$.  Varying the relative strength of the cooling $\kappa$ does not change the unstable effective temperature range, but does vary the corresponding stresses; a stronger cooling is simply overcome at a higher stress.

While I have shown there are instabilities, I have not proven they are where the effective temperature as a function of stress is multivalued.  For the $\alpha > 1$, $\Gamma_0$ is a monotonically increasing function of the applied stress $s_0$, i.e. $\pd{\Gamma_0}{s_0} > 0$.  Therefore, the section of the curve where $\pd{s_0}{\chi_0} = \pd{\Gamma_0}{\chi_0} \deriv{s_0}{\Gamma_0} < 0$ requires $\pd{\Gamma_0}{\chi_0} < 0$.  Setting the LHS of Eq.~\eqref{eq:15} to 0 and differentiating with respect to $\chi_0$ yields
\begin{equation}
  \label{eq:24}
  \pd{\Gamma_0}{\chi_0} \propto \frac{ 1 - \beta }{ \chi_0^2 } \left(\chi_0 - \frac{T}{T_Z}\right) - \frac{ \chi_\infty - \frac{T}{T_Z} }{ \chi_\infty - \chi_0 } \propto \omega(0).
\end{equation}
Therefore, as one would expect, the section of the steady-state $\chi$ vs.\ $s$ curve where $\pd{s_0}{\chi_0} < 0$ is unstable.

\begin{figure}[tbp]
  \centering
  \includegraphics[width=\columnwidth,clip]{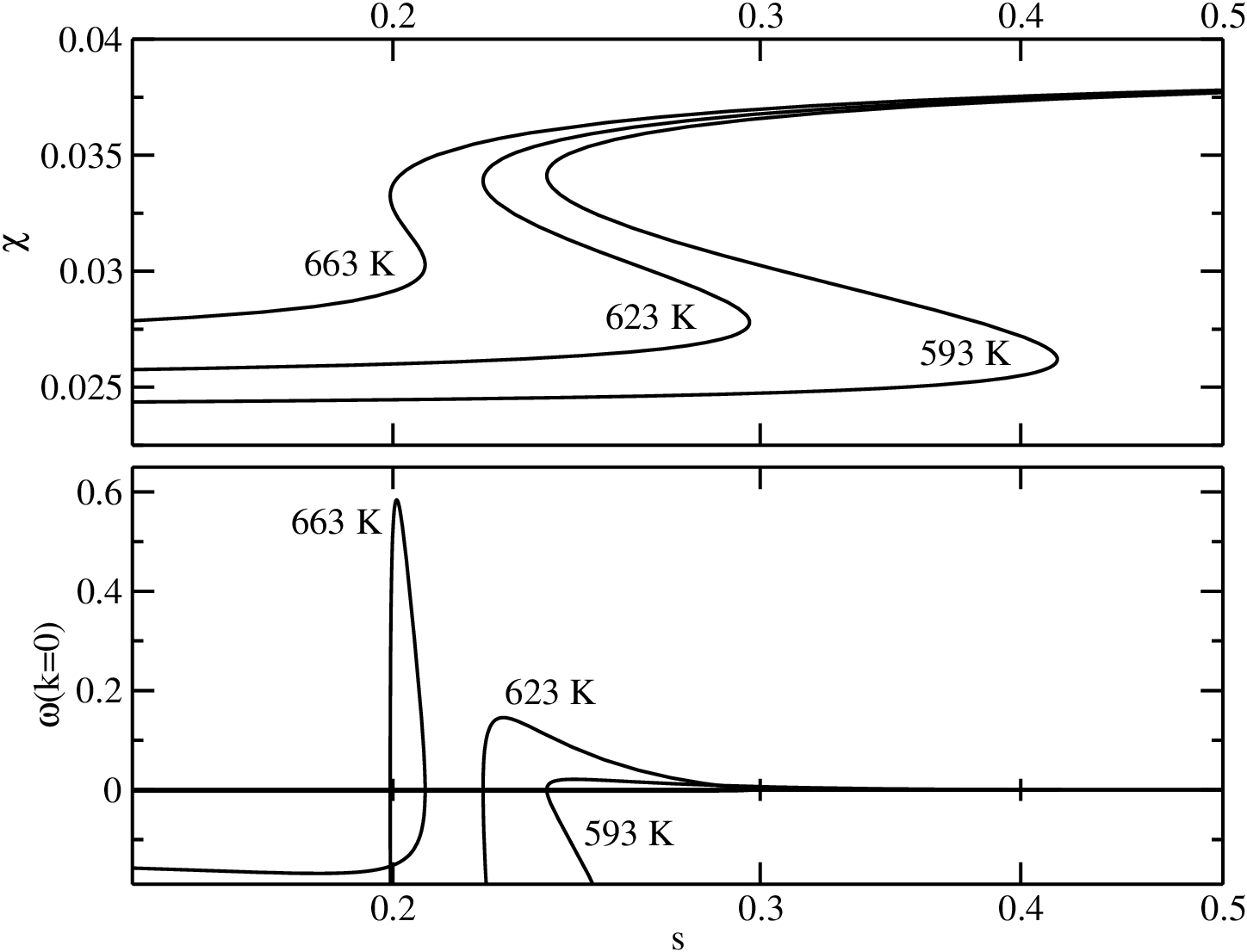}
  \caption[Effective temperature $\chi$ and instability exponent $\omega(0)$ vs.\ stress]{Effective temperature $\chi$ and instability exponent $\omega(k=0)$ (in units of $\frac{\varepsilon_0}{\tau_0 c_0}$) vs.\ stress $s$ for the steady state solution of the STZ equations at varying temperatures for $\beta = 0.6$, $\kappa = 10^{-6}$, $T_z = 24500 K$, $T_R = 850 K$, $T_0 = 836 K$, $T_1 = 45 K$, $T_\infty = 937 K$}
  \label{fig:stability_graph}
\end{figure}

Figure~\ref{fig:stability_graph} shows the effective temperature and instability for three different temperatures.  A comparison of the graphs of $\chi$ and $\omega(k=0)$ shows that, as predicted, the instability occurs at those stresses where the $\chi$ is multivalued.

  The system is only unstable for a finite range of strain rates; a material deforming sufficiently fast will be homogeneous.  Experimental evidence for such a transition back to homogeneous flow has been seen in some metallic glasses\cite{Schuh2004}.  When the material is driven at an unstable strain rate, it can break up into two or more regions of different strain rates.  Because the dispersion relation, Eq.~\eqref{eq:18}, does not have a peak at nonzero $k$, there is no length scale describing the width or spacing of the shear bands.  The only length scale is the width $a$ of the boundary between the two regions
\begin{multline}
  \label{eq:25}
   a^2 = D_\chi \left/ \left\{ \kappa \rho e^{-\beta/\chi_0} \vphantom{\left[ \frac{1-\beta}{\chi_0^2} \left(\chi_0 - \frac{T}{T_Z}\right) - \frac{ \chi_\infty - \frac{T}{T_Z} }{ \chi_\infty - \chi_0 } \right]} \right. \right.\\
   \left. \left. \times \left[ \frac{1-\beta}{\chi_0^2} \left(\chi_0 - \frac{T}{T_Z}\right) - \frac{ \chi_\infty - \frac{T}{T_Z} }{ \chi_\infty - \chi_0 } \right] \right\} \right. .
\end{multline}

\subsection{Comparison to Experiment}
\label{sec:comp-exper}

Before doing quantitative fits to the experiments, we must be explicit in mapping the STZ equations as written to the results of a 3-D uniaxial compression test, for which we will follow the procedure laid out in \cite{Langer2004}.
  The stress tensor in such tests is, denoting the axis of compression as $x$, $\sigma_{xx} = \sigma'$ and all other components zero.  By using the second invariant of the deviatoric stress tensor to define the STZ equation stress $s$ such that $s = \frac{1}{\sqrt{3}} \frac{\sigma'}{\bar{\mu}}$, the STZ yield stress becomes qualitatively similar to the von Mises yield criterion.
  Likewise, the scalar $m$ will correspond to the second invariant of an STZ bias tensor.  With these conventions, Eq.~\eqref{eq:3} describes $\dot{\varepsilon}_{xx}$, up to a factor of $\sqrt{\frac{4}{3}}$ which can be absorbed into $\varepsilon_0$, while other equations remain unchanged.

As shown in \cite{Langer2004}, at low stresses, the material will approximate a Newtonian liquid with a measured viscosity $\eta_N(T)$ given by
\begin{equation}
  \label{eq:27}
  \eta_N(T) = \eta_0 \frac{1}{\alpha(T)} \exp\!\left( \frac{T_z}{T} + \alpha(T) \right),
\end{equation}
where $\eta_0 = \frac{\sqrt{3} \bar{\mu}}{\varepsilon_0 / \tau_0}$.  The metallic glass viscosities are separated into two groups.  First the high temperature limit of Eq.~\eqref{eq:27}, $\eta_N(T) = \frac{2 \eta_0}{T_R} T \exp\!\left(\frac{T_Z+\frac{T_R}{2}}{T}\right)$ is used to fit the higher temperature values\cite{AMRB99}; then those constraints are used when the lower temperature values\cite{JLGR03} are fit with the full expression.  The best-fit parameters are $\eta_0 = 1.6 \times 10^{-11}$~Pa~sec, $T_0=836$~K, $T_R = 850$~K, and $T_1 = 45$~K.  $T_Z$ is $24{,}500$~K which is on the order of the value of $\mu l^3$ for an STZ of a few atoms in diameter.
  The fit was done more rigorously than and improves upon the previous\cite{Langer2004}: only two of the low temperature viscosities are not well-explained by Eq.~\eqref{eq:27}, the first and the fourth.  Unlike \cite{Langer2004}, I will ignore any possible errors in the temperatures reported in the experiment, and use the reported values.

\begin{figure}[tbp]
  \centering
  \includegraphics[width=\columnwidth,clip]{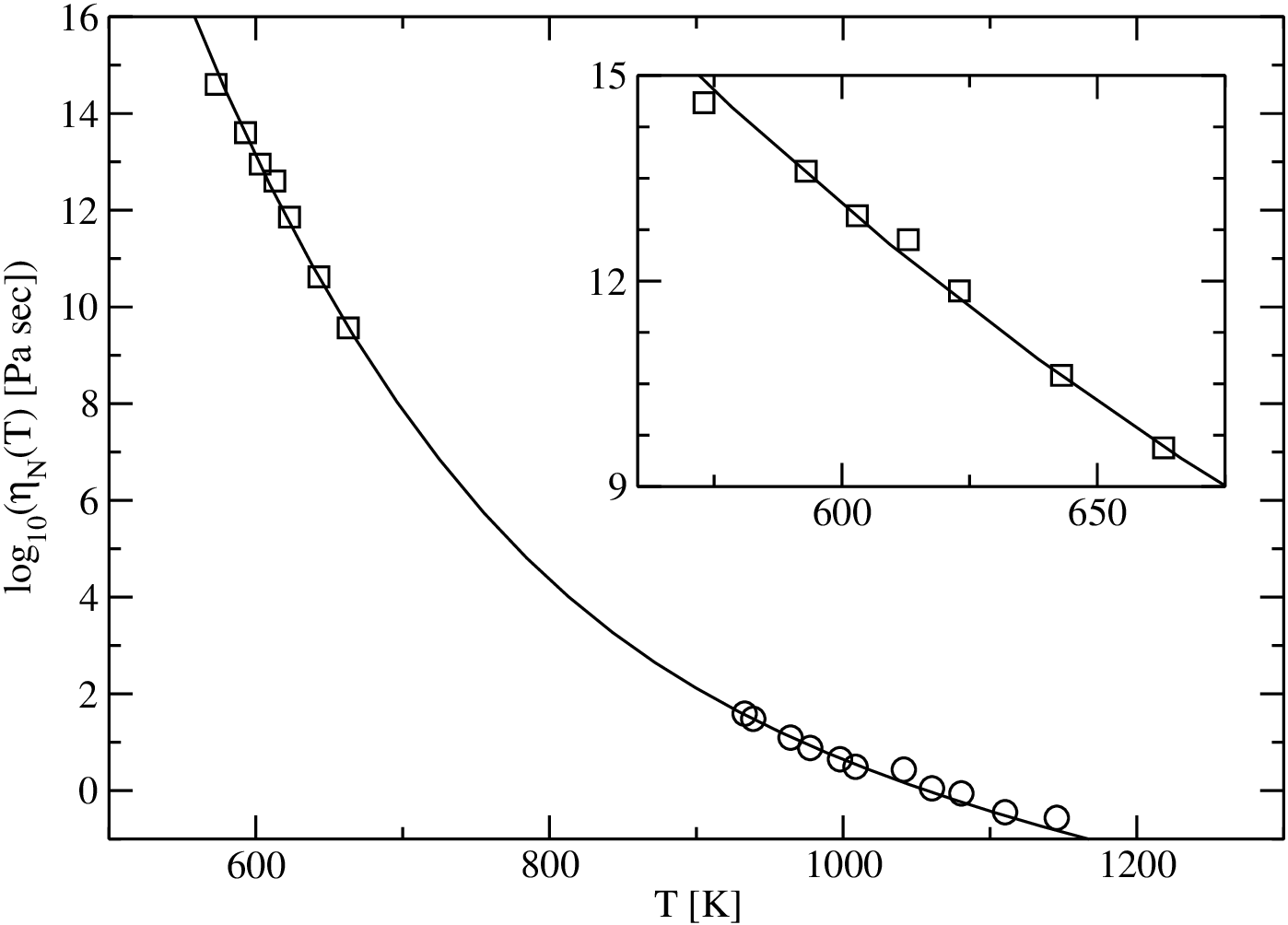}
  
  \caption[Analytic fit of metallic glass viscosity to the predicted curve from the STZ model with effective temperature]{Analytic fit of metallic glass viscosity to the curve predicted from the STZ model with effective temperature, Eq.~\eqref{eq:27}, using the Cohen-Grest expression for $\alpha(T)$, Eq.~\eqref{eq:11}, with $\eta_0 = 1.6 \times 10^{-11}$~Pa~sec, $T_z = 24{,}500$~K, $T_0=836$~K, $T_R = 850$~K, and $T_1 = 45$~K.  The circles and squares are experimental results for Vitreloy 1 from \cite{AMRB99} and \cite{JLGR03} respectively.}
  \label{fig:viscosity}
%
%
\end{figure}

Before fitting the uniform steady-state behavior at higher strain rates, it will be useful to reduce the parameter space by estimating some of the parameters beforehand.  If we assume $T_\infty = T_0$ as assumed previously, then to match the experimental observation of localization at $T = 663$ K, $\beta$ must be less than 0.47.  I feel that is too low and instead shall use $T_\infty = T_{solidus}$ which is 937~K for the metallic glass\cite{Peker1993}, in which case $\beta < 0.63$.  I shall stay at the high end of that limit and use $\beta = 0.6$.

Next, for $\bar{\mu}$, which is assumed to be the yield stress, I propose the following argument: Imagine applying a constant strain rate to a material initially at rest.  The effective temperature dynamics are slow enough that the number of zones is essential unchanged during the initial stress rise.  The peak stress is determined by the stress necessary to produce the applied strain rate purely plastically.  If the strain rate is small, this can be accomplished purely from plastic creep; the stress will be below the yield stress and highly dependent on temperature.  If the strain rate is large, the system will have to enter the regime of STZ flow; the stress will be above the yield stress and less dependent on temperature.  This exact effect can be seen in the metallic glass data in Fig.~4 in~\cite{JLGR03}, with a distinct change in the temperature dependence of those experiments where the peak stress was above 1500~MPa.  In the 2-D simulations, this corresponds to $\bar{\mu} = 500 \sqrt{3}$~MPa.

\begin{figure}[tbp]
  \centering
  \includegraphics[width=\columnwidth,clip]{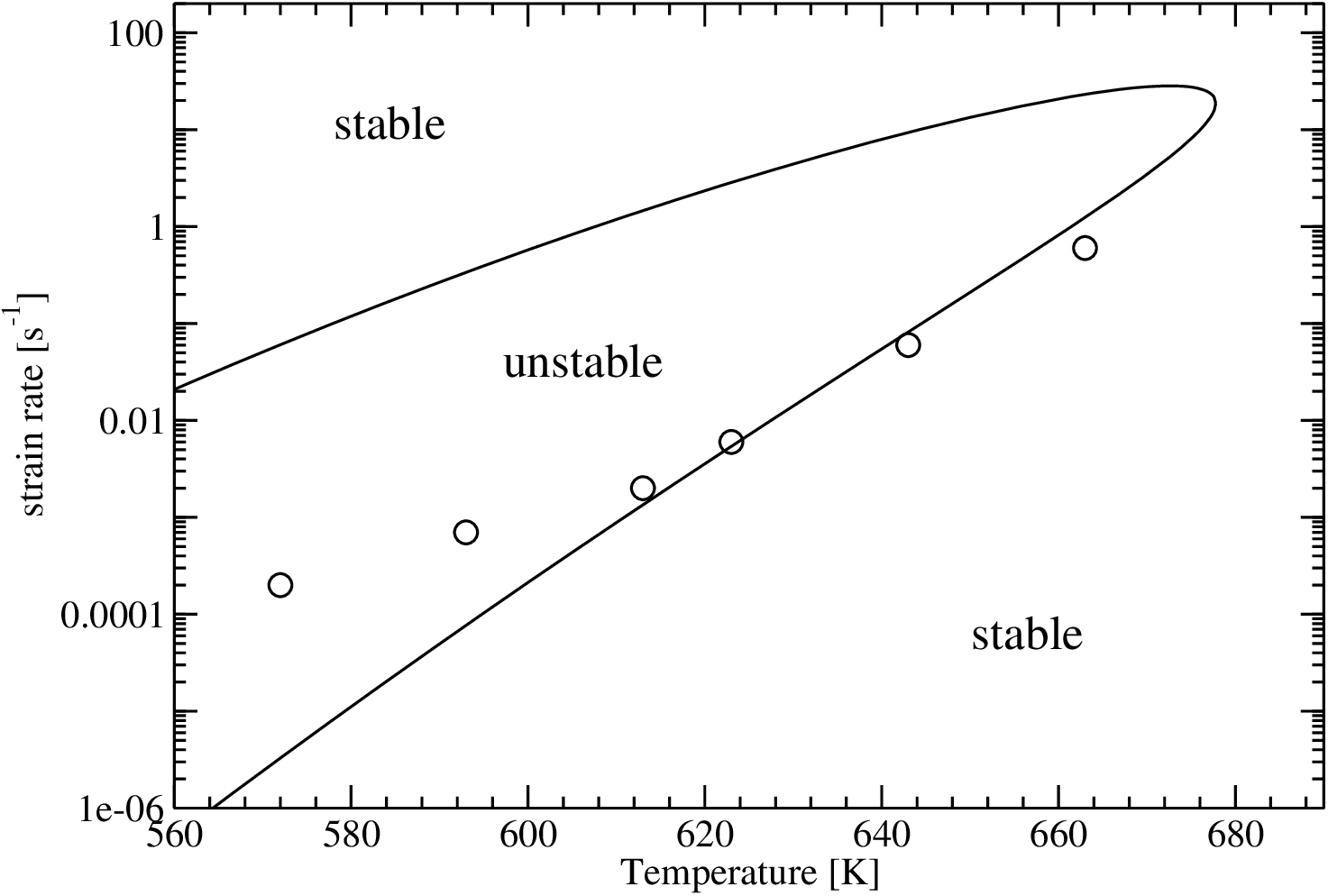}

  \caption[Stability map of the uniform steady-state solutions of the STZ model with effective temperature]{Stability map of the uniform steady-state solutions of the STZ model with effective temperature for $T_\infty = T_{solidus} = 937$~K, $\beta=0.6$, $\kappa = 2.5 \times 10^{-6}$, $\bar{\mu} = 500 \sqrt{3}$~MPa.  The other parameters are the same as in Fig.~\ref{fig:viscosity}.  Strain rates inside the curve are unstable.  The circles are experimental data for Vitreloy 1 taken from Fig.~11 in \cite{JLGR03}.}

  \label{fig:localization_strain_rate_vs_T}
\end{figure}

That leaves $\kappa$.  Instead of trying to fit uniform steady-state stress vs strain rate for all eight temperatures directly, I used the minimum strain rate at localization as a function of temperature first.  This fit is shown in Fig.~\ref{fig:localization_strain_rate_vs_T}.  Because of the large error bars, and the vagueness in deciding whether a sample is localized or not, the fit was done by eye to yield a value of $2.5 \times 10^{-6}$.  While the model agrees well at high temperatures, at lower temperatures it underestimates the strain rate at localization by a factor of 100.  Remember, the predictions are for the steady state results. Later, when the dynamics are included, the difference will be apparent.

\begin{figure}[tbp]
  \centering
    \includegraphics[width=\columnwidth,clip]{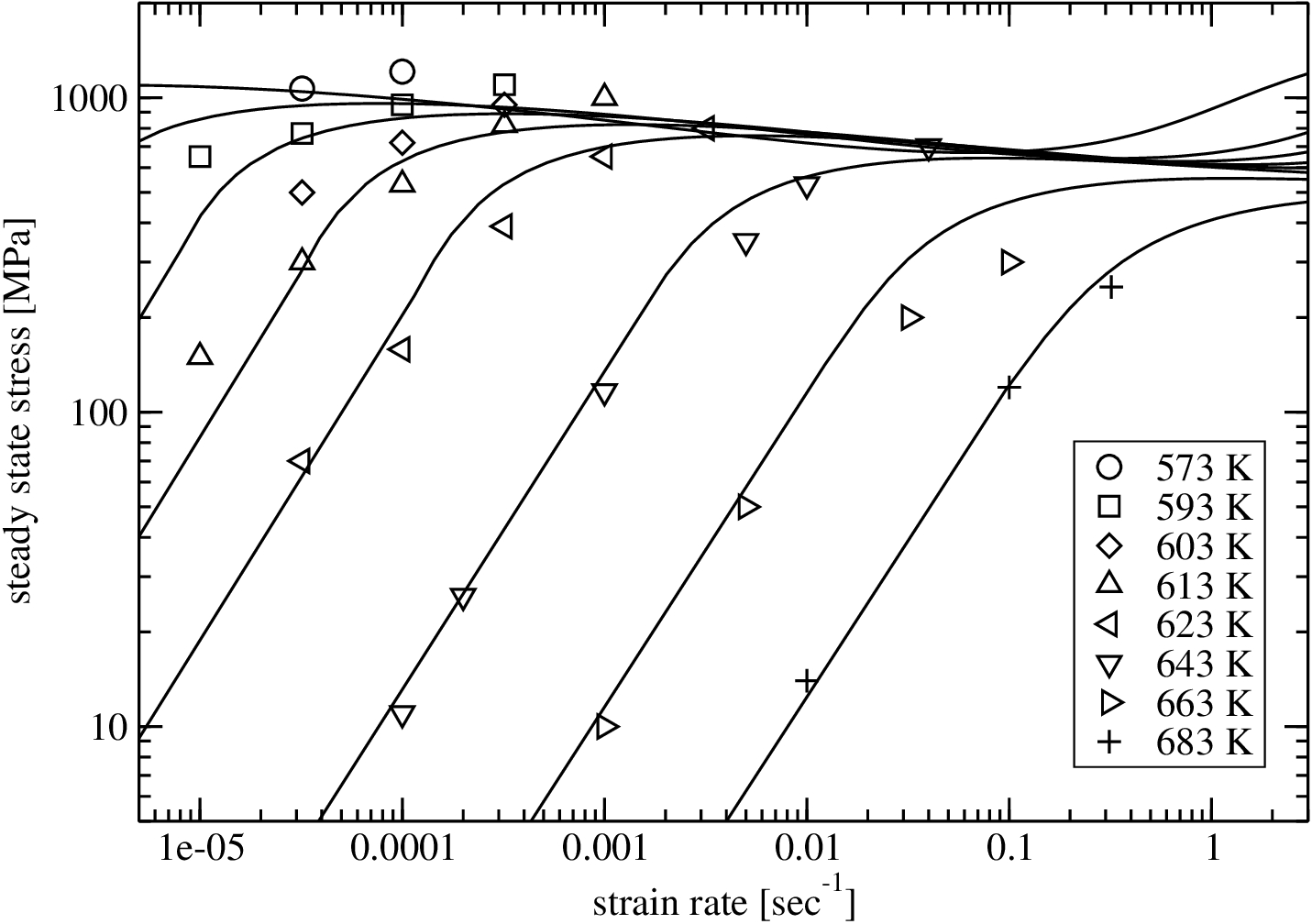}

  \caption[Steady-state stress vs.\ strain for uniform solutions of the STZ model with effective temperature]{Steady-state stress vs.\ strain rate for uniform solutions of the STZ model with effective temperature using the same parameters as Fig.~\ref{fig:localization_strain_rate_vs_T}.  The symbols are the experimental results for Vitreloy 1 from~\cite{JLGR03}}
  \label{fig:stress_vs_strain_rate_fits}
\end{figure}

Figure~\ref{fig:stress_vs_strain_rate_fits} compares the steady-state stresses vs.\ strain rates predicted by the model to those reported in the experiment\cite{JLGR03}.  Compared to the fit in \cite{Langer2004} the high temperature, low stress data part of the fit is improved, but the low temperature, high stress parts are worse.

\section{Shear Localization: Nonlinear Stability}
\label{sec:shear-local-dynam}

\subsection{Strain Rate Dependence}
\label{sec:steady-strain-rate}

Now that we've found the parameter ranges with the best prospects for banding, I will numerically solve the system of equations in the linearly unstable regime and study the full results.
I will model a material under a controlled constant total strain rate, first starting with a system at an unstable steady state solution of the dynamics.  Then I will examine systems starting at rest, and the effects the transient response has on the stability.  Finally I will vary the temperature and study the effects of the initial STZ density.  I shall assume the elastic wave speed is great enough that the stress can be assumed to be uniform at all times.  The uniform stress forces the STZ bias ratio $m$ to also be uniform.  The time evolution of the stress is determined by
\begin{equation}
  \label{eq:29}
  \dot{s} = 2 \mu \left( \dot{\varepsilon}^{tot} - \frac{1}{L} \int_0^L \dot{\varepsilon}^{pl}(x) dx \right).
\end{equation}
This equation with Eqs.~\eqref{eq:8} and \eqref{eq:16} completely describes the dynamics of the system.

The boundary size of any shear bands can be estimated from Eq.~\eqref{eq:25} and an estimate of the diffusion constant.   According to Ono et al.~\cite{Ono2002}, despite being non-equilibrium, the diffusion constant in the direction perpendicular to the flow still obeys the Stokes-Einstein relation $D \propto \frac{T_{eff}}{l \eta}$  The zone size $l$ is on the order of a few atoms; $l \sim 10^{-9}$~m.  Despite no longer being in a viscous regime, the viscosity can be approximated by the average viscosity, $\frac{\sigma}{\dot{\varepsilon}}$.  Since the effective temperature and the strain rate vary greatly between the cold side, where $\chi = \chi_-$, and the hot side, where $\chi = \chi_+$, the diffusion constant varies from $10^{-25}$ to $10^{-19}$~$\mathrm{m}^2/\mathrm{sec}$.  I shall use a value at the high end of this range, $D = 10^{-20}$~$\mathrm{m}^2/\mathrm{sec}$, which is in rough agreement with the experimentally measured atomic diffusion constants.\cite{MMVN+01, Geyer1995}  I shall also ignore any temperature dependence.  Any errors in this approximation should not seriously affect the results; as this is the only length scale in the problem, all the other lengths can be rescaled to compensate.

Using this value in Eq.~\eqref{eq:25} gives an estimate of the boundary width on the order of a few angstroms.  To avoid the divergence from the term in brackets at $\chi_\pm$, the estimate $\chi_0=\frac{1}{2} \left( \chi_+ + \chi_- \right)$ was used.
  It should be noted that although, over the temperature of the experiment, the strain rate and diffusion constant vary by three orders of magnitude, the boundary size stays relatively constant as the temperature varies between 577 and 663~K.  This can be seen from rewriting Eq.~\eqref{eq:25} (ignoring the slowly varying, non-exponential term in brackets) as
\begin{equation}
  \label{eq:30}
  a^2 \sim \frac{D_\chi}{\kappa \rho e^{-\beta/\chi}} \propto \frac{\chi \dot{\varepsilon}}{\sigma} \frac{1}{e^{-1/\chi} \Gamma} \propto \frac{ \chi }{ \sigma^2 }.
\end{equation}
The number of zones, $\exp(-1/\chi)$, is the most varying factor as the temperature decreases, but since it affects both the plastic work and the Stokes-Einstein approximation of the diffusion constant in identical ways the boundary size does not vary by more than a factor of 10.

The numerical setup is as follows: The systems start in a uniform state given by $s_0$, $m_0$, $\chi_0$, and have periodic boundary conditions.  All systems are $0.1$ microns in size and divided into 1000 points.  Inhomogeneities are introduced by adding a random noise, $\Delta \chi \left( x_i \right)$, to the effective temperature at each point.  These random variables are independent and identically distributed uniformly over the range $\{-0.5\% \chi_0,0.5\% \chi_0\}$.  The system is then evolved according to the equations of motion up to strains of $\varepsilon_f=0.4$.  From this, a final, steady-state, stationary profile is also computed by minimizing the time-derivatives of the state variables as expressed in Eqs.~\eqref{eq:8}, \eqref{eq:16}, and \eqref{eq:29}, starting with the results of the simulation of the dynamics at strain $\varepsilon_f$.  The same material parameters as in Fig.~\ref{fig:localization_strain_rate_vs_T} are used with $\mu$ having the experimental value of 35~GPa, and $\varepsilon_0 = \exp(1/\chi_\infty)$.  The remaining parameter $c_0$ should be of order 1 and, by comparison of the response of the material as a shear band forms, and the experimental results, a value of $c_0=0.6$ is used.\cite{Foglia2006}

\begin{figure}[tbp]
  \centering
  \includegraphics[width=\columnwidth,clip]{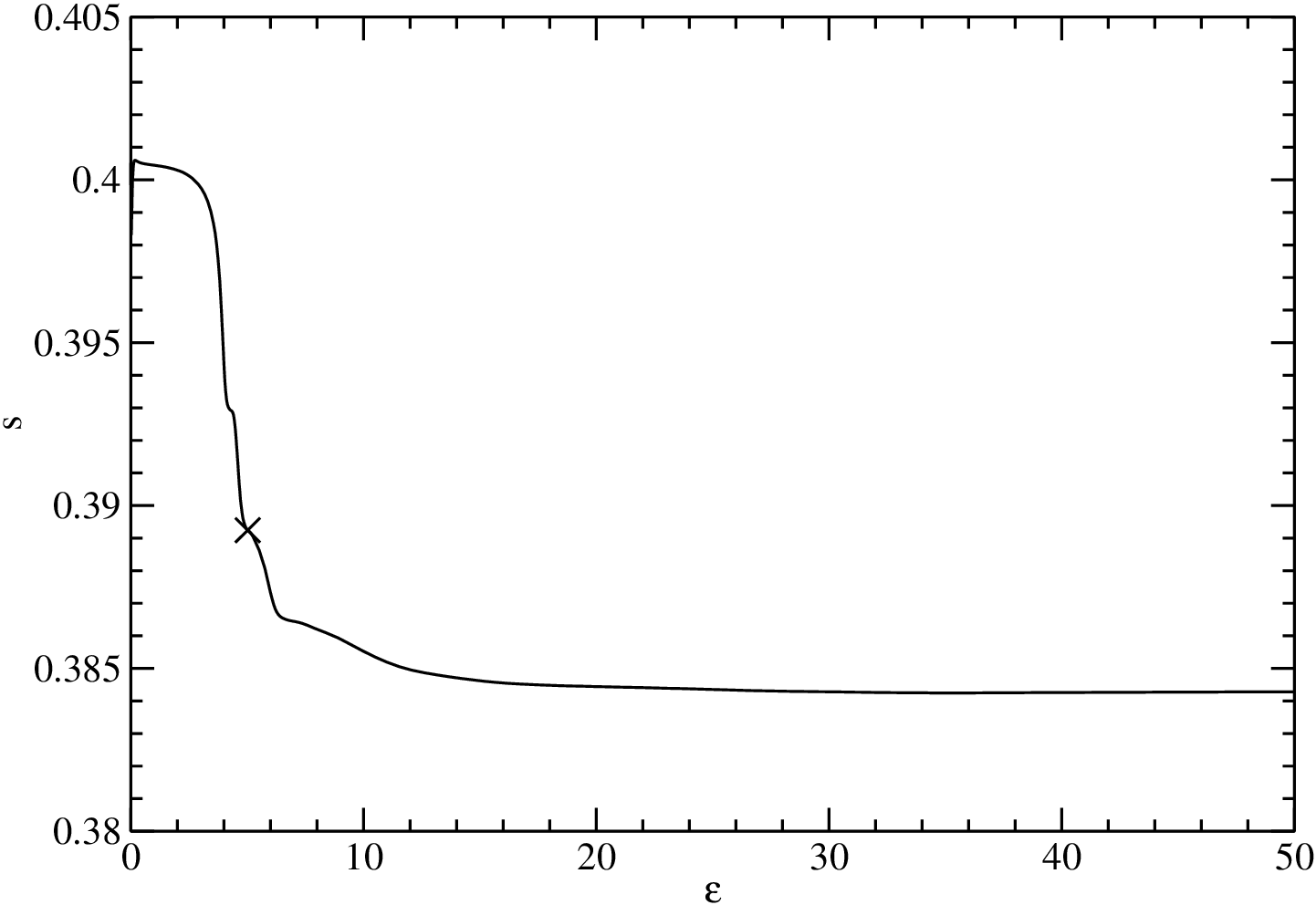}
  \caption[Stress vs.\ strain for a system initially in an unstable uniform steady-state as a shear band forms]{The stress relaxation of an STZ material as a shear band forms.  The system is started at a uniform unstable point, with a small amount of noise added to the effective temperature profile, then evolves at a fixed strain rate of $1$~$\mathrm{sec}^{-1}$, according to Eqs.~\eqref{eq:8}, \eqref{eq:16}, and \eqref{eq:29}.  The ``X'' represents the point at which localization is defined to occur.}
  \label{fig:s_vs_strain:most_unstable}
\end{figure}

  Figure~\ref{fig:s_vs_strain:most_unstable} shows the stress relaxation as a function of strain for a system with the previously listed parameters, at a bath temperature of 643 K, starting at the unstable uniform solution of $\chi_0 = \frac{1}{2} \left( \chi_+ + \chi_- \right) = 0.313$, $s_0 = 0.4$, $m_0 = 0.98$.  This corresponds to an applied strain rate of $1$ $\mathrm{sec}^{-1}$.

Localization is determined by the Gini coefficient of the strain rate distribution.  The Gini coefficient can be defined as the average difference between all possible pairs of values (including pairs where both values are chosen to be the same point) divided by the average and normalized to be 1 in the case of an infinite system where only one value is nonzero.  Denoting the Gini coefficient by $\phi$, the value is defined by\cite{MathWorld-Gini}
  \begin{equation}
    \label{eq:gini_coefficient}
    \phi = \frac{1}{2 N^2 \overline{\dot{\varepsilon}^{pl}}} \sum_i \sum_j \left| \dot{\varepsilon}^{pl}(x_i) - \dot{\varepsilon}^{pl}(x_j) \right|,
  \end{equation}
where $N$ is the number of sites, and $\overline{\dot{\varepsilon}^{pl}}$ is the average of the plastic strain rate distribution.  $\phi$ ranges from 0 for a completely uniform distribution to 1 for a system where all the strain is occurring at one infinitesimally small point.  Any systems with $\phi > 0.5$ will be considered localized.  This value is arbitrary, but easily measurable, and in most of the results $\phi$ is sharply increasing when this criterion is surpassed.

In Fig.~\ref{fig:s_vs_strain:most_unstable}, the point where the material localizes is marked with an ``X''.  Any results beyond that point should be taken with a grain of salt, both because the experiments cease at that point so there is nothing with which to compare, but also because the sharp sudden onset of a shear band could lead to fracture.

The effective temperature profile shows five shear bands of high strain rate and effective temperature.  There was no a priori reason to expect five bands; with no relevant length scale, the number of zones is determined by the initial noise.  Similar runs were done for other noise realizations, and the number of bands varied from 2 to 5.  In comparison to experiments though, the shear bands computed are much thinner than those reported.  In a similar metallic glass, band widths of 50 nm were reported\cite{Yang2004} and are possibly higher in this particular metallic glass.\cite{Wright2001}  The dashed line is the strain rate, which because of its strong dependence on $\chi$ varies even more sharply than the effective temperature.

\begin{figure}[tbp]
  \centering
  \includegraphics[width=\columnwidth,clip]{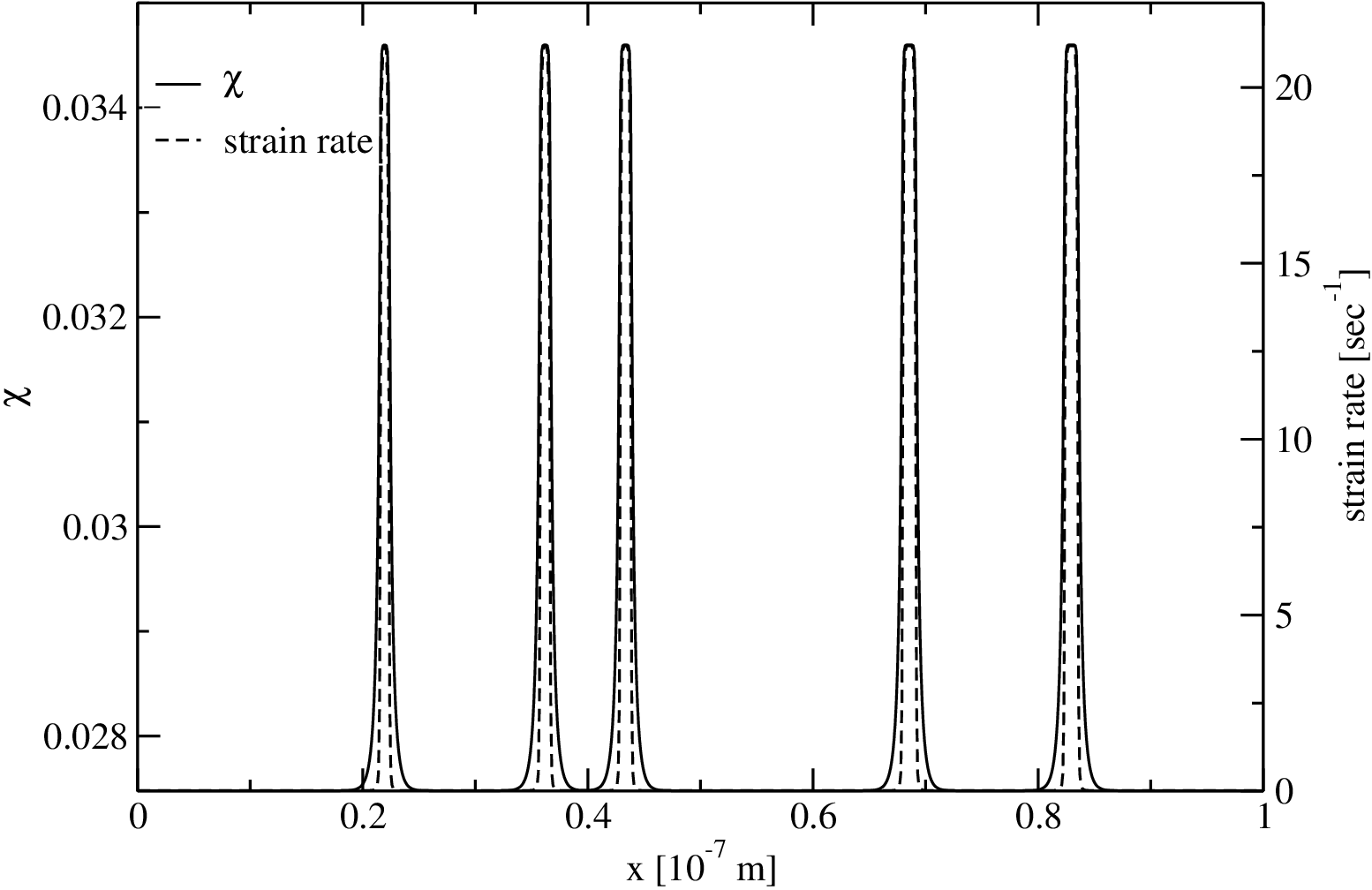}
  \caption[Steady-state profile of the effective temperature of the effective temperature $\chi$ and strain rate vs.\ position for a system driven at an unstable strain rate]{Steady-state profile of the effective temperature $\chi$ and strain rate vs.\ position $x$ of the controlled-strain-rate driven system of Fig.~\ref{fig:s_vs_strain:most_unstable}.  The lengths are in units of the system size, $10^{-7}$ m.}
  \label{fig:chi_profile:most_unstable}
\end{figure}

These results are interesting and prove an instability is possible, but they do not directly correspond to experiments, the experiments are not started at a uniform shear solution.  Instead they are started with no stress ($s_0 = 0$) and no bias in the STZ population ($m_0 = 0$).  The stability of the initial, transient response may differ both qualitatively and quantitatively from the stability of the uniform steady-state solution.  In an examination of a Spaepen's free volume model\cite{Spaepen1977}, Huang, et al.\cite{Huang2002} showed that although the uniform steady state was stationary, an instability can arise during the initial response during a constant-strain-rate experiment.

To study the effects of the initial loading, four additional simulations are run.  Like the previous, a system is created with a width of $0.1$ microns discretized into 1000 sites.  Unlike before the initial stress and bias are zero ($s_0 = 0$, $m_0 = 0$).  The initial noise of the effective temperature is again $0.5\% \chi_0$, but $\chi_0$ is equilibrated to the bath temperature which again is chosen to be 643~K.  This system is then used as the starting configuration for four controlled strain rate tests, one at each of the strain rates in Fig.~2 from \cite{JLGR03}.

\begin{figure}[tbp]
  \centering
  \includegraphics[width=\columnwidth,clip]{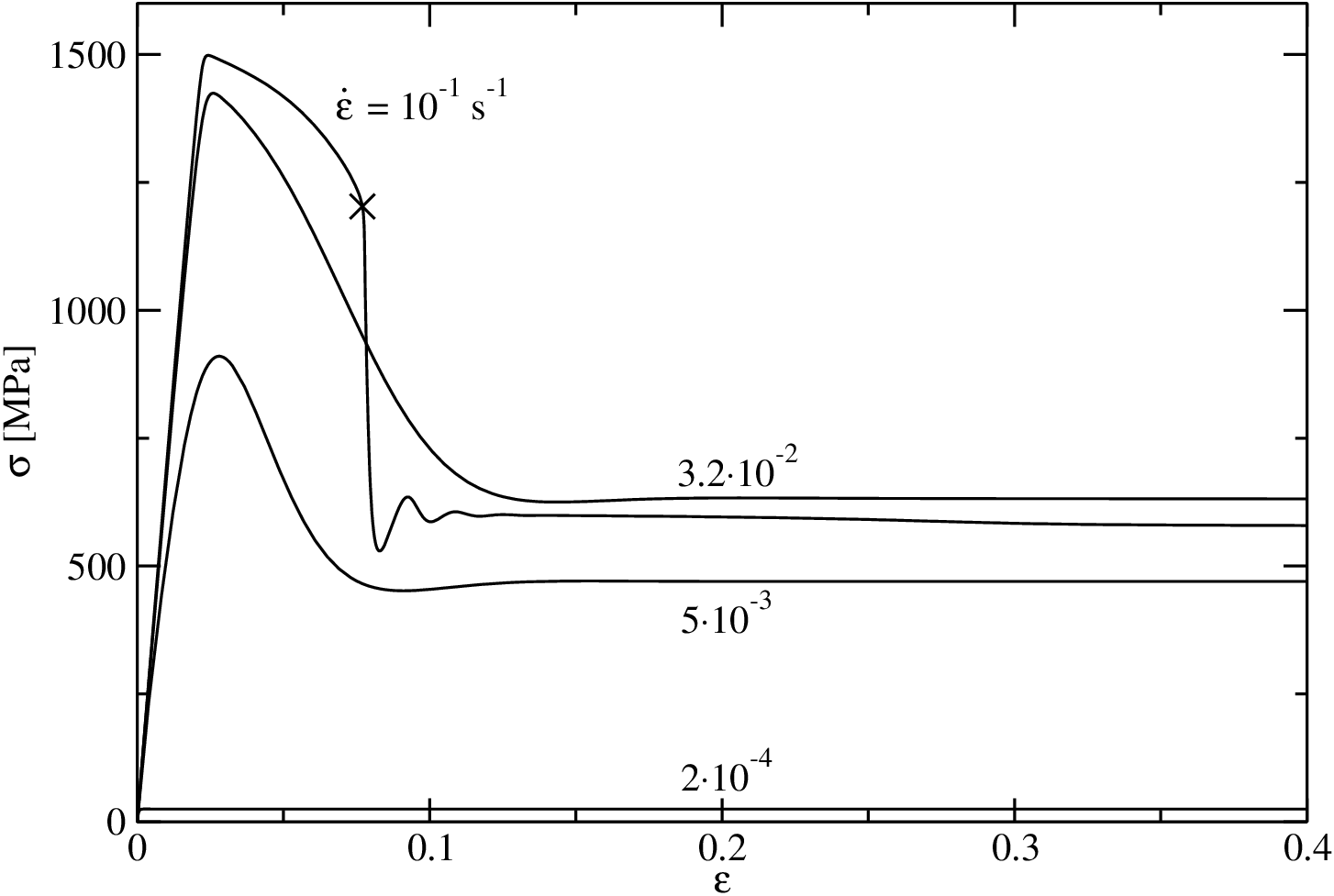}
  \caption[Stress vs.\ strain for a system initially at rest for varying strain rates]{Stress vs.\ strain for a system initially at rest for four different strain rates at $T=643$~K.  In addition to the parameters of figure~\ref{fig:localization_strain_rate_vs_T}, $c_0 = 0.6$, $\varepsilon_0 = e^{1/\chi_\infty}$, $D_\chi = 10^{-20} \mathrm{m}^2/\mathrm{s} \left(\frac{\varepsilon_0}{\tau_0 c_0}\right)^{-1} = 1.12 * 10^{-40}$, and the number of lattice points is 1000.}
  \label{fig:stress_vs_strain}
\end{figure}

The stress as a function of strain for these simulations is plotted in Fig.~\ref{fig:stress_vs_strain}.  Like the metallic glass experiment, the lowest three systems stay homogeneous throughout the run, while the highest strain rate localizes suddenly.  The graphs for the two highest of non-localizing strain rates, $3.2 \times 10^{-2}$ and $5.0 \times 10^{-3}$~$\mathrm{sec}^{-1}$, display a stress overshoot, also like the experimental data.

\begin{figure}[tbp]
  \centering
  \includegraphics[width=\columnwidth,clip]{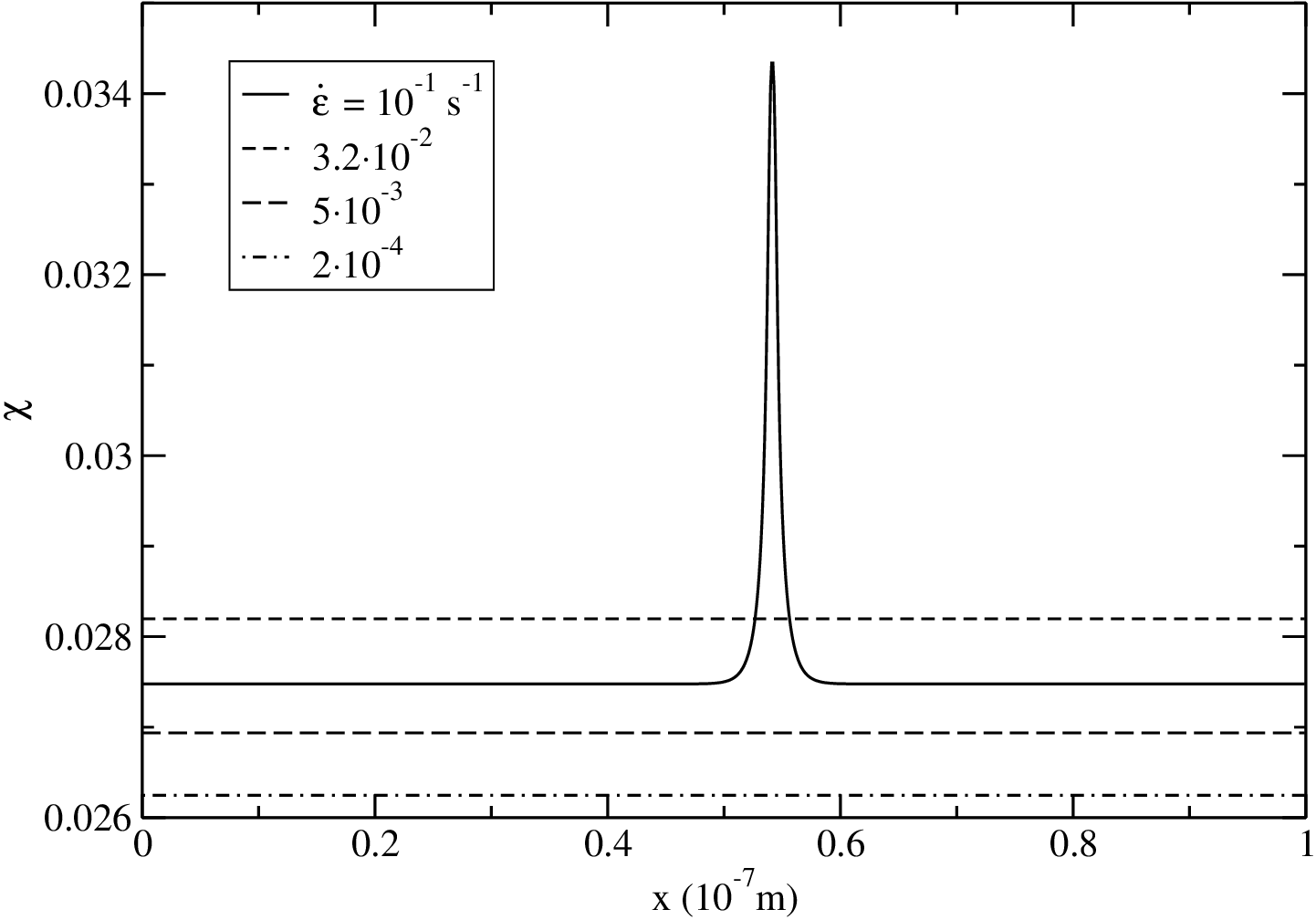}
  \caption[Steady-state effective-temperature profiles for systems initially at rest for varying strain rates]{Steady-state effective-temperature profiles of the systems in Fig.~\ref{fig:stress_vs_strain}.}
  \label{fig:steady_state_profiles}
\end{figure}

  The highest strain rate results also shows a stress overshoot quickly followed by the onset of localization, both in the numerical results and in the actual experiment.  Final steady-state effective temperature profiles are computed by the same procedure as previously explained and plotted in Fig.~\ref{fig:steady_state_profiles}.  Note how the system at the higher strain rate can find a lower average effective temperature by localizing the shear.  The strain rate is so much larger in the shear band, the remainder of the material can relax to a lower effective temperature.

Unlike the system that started in the uniform steady state, the time it takes for the highest strain rate system in Fig.~\ref{fig:stress_vs_strain} to shear localize is much shorter, less than 1 second compared to the 5 seconds for a system at the most unstable strain rate.  The speed is even more dramatic when considering that the dominating time scale is the strain rate, which is 10 times faster in the earlier case.

\begin{figure}[tbp]
  \centering
  \includegraphics[width=\columnwidth,clip]{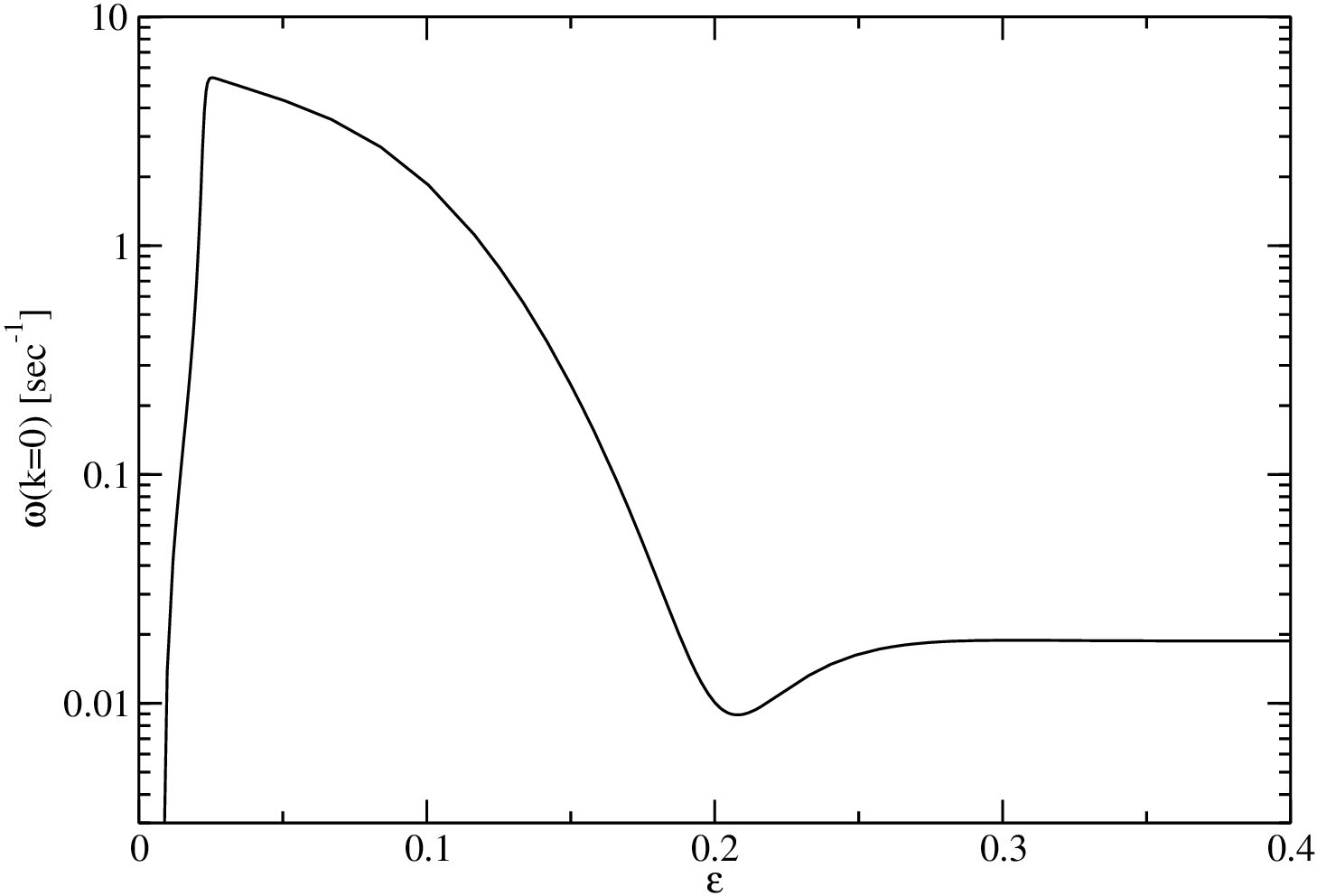}
  \caption[Stability exponent of the uniform solution of the STZ equations as a function of strain for a material initially at rest]{Stability exponent $\omega(0)$ given by Eq.~\eqref{eq:18}, for the uniform solution of the STZ equations as a function of strain for  a material initially at rest and under an applied strain rate of $\dot{\varepsilon} = 10^{-1} \mathrm{s}^{-1}$, the highest strain rate in Fig.~\ref{fig:stress_vs_strain}.}
  \label{fig:stability_exponent_of_uniform_transitory_solution}
\end{figure}

A similar localization-during-transient response was seen by Huang, et al.\cite{Huang2002} during their numerical analysis of Spaepen's free-volume model of plasticity\cite{Spaepen1977}.  Although that model has only stable uniform steady states, they showed that the transient response of such a system could lead to a localization.  (Their localization was only temporary, but that need not the case with the STZ model with effective temperature.)  Following their logic, a time-dependent stability exponent can be calculated by examining the stability of the uniform solution as the latter evolves in time\cite{Bai1982}.  Instead of perturbing around a time-independent solution, the perturbation is done with respect to a time-dependent uniform solution $s_0(t)$, $m_0(t)$, and $\chi_0(t)$.  The result is exactly the same as Eq.~\eqref{eq:17}, but no longer can the stress-dependence be explicitly removed as in  Eq.~\eqref{eq:18}.  Figure~\ref{fig:stability_exponent_of_uniform_transitory_solution} shows the largest stability exponent $\omega(0)$ taken from Eq.~\eqref{eq:17} as a function of strain rate for a uniform system with the same parameters as in Fig.~\ref{fig:stress_vs_strain} at a strain rate of $10^{-1}$  $\mathrm{s}^{-1}$.  Because of the initial stress rise, $\Gamma$ becomes very large near the stress peak and the solution becomes very unstable to localization.  In comparison to the final stability exponent of the uniform solution, the peak value is over two orders of magnitude larger.

\subsection{Temperature Dependence}
\label{sec:varying-temperature}

Now that we've examined the effects of varying the strain rate, we will vary the bath temperature as well.  To do so, first more consideration must be made as to the initial values of the effective temperature.  So far, we have examined the model for cases where the initial effective temperature is equal to the bath temperature.  If this were always the cases, there would be no reason that the effective temperature had to be ``effective'' and couldn't actually be a traditional thermal temperature.  In which case, all the results seen so far would simply be examples of adiabatic shear bands.  The only difference would be the diffusion constant, but that only sets the length scale.  But as the initial temperature lowers, that equality will no longer hold.  Instead we must estimate the initial effective temperature from the thermal history of the material.

Consider a material that is first cooled from the melting temperature to room temperature, then heated to and annealed at a temperature at which it will be tested.  Initially, $T_{eff} = T$, but as $T$ drops, the cooling dynamics become slower and slower, and eventually the effective temperature of the system falls out of thermal equilibrium at some value.  Assuming  a final bath temperature of 295 K and a cooling rate of 1 K/s (the typical cooling rates in the production of Vitreloy 1 are between 10 and 0.9 K/s \cite{Peker1993, Kim1994}), and using the same parameters as in the previous section, the effective temperature plateaus at 640 K after roughly 20 minutes of cooling and stays frozen at that value for much, much longer than laboratory timescales (over $10^{10}$ years).

\begin{figure}
  \centering
  \includegraphics[width=\columnwidth,clip]{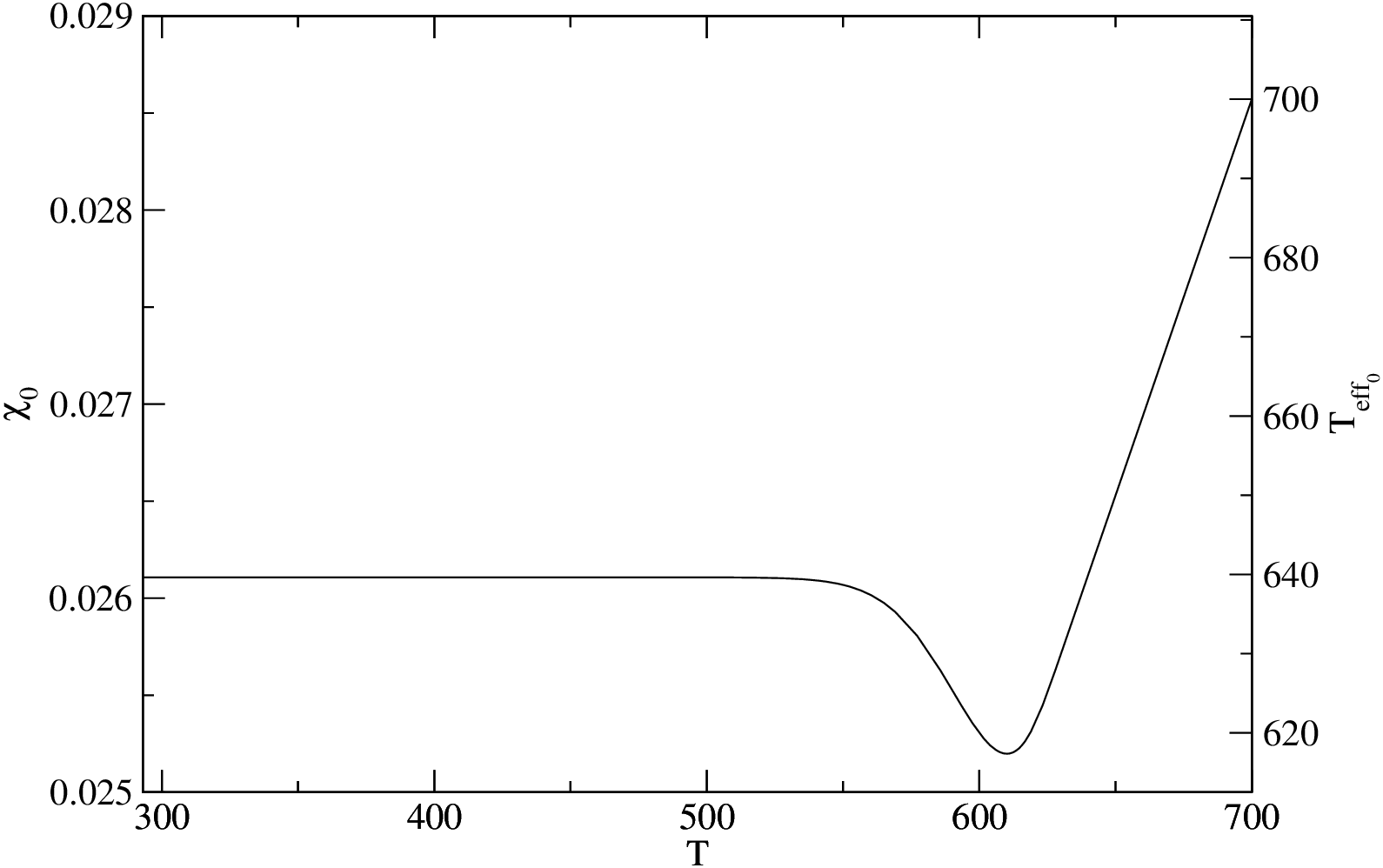}
  \caption[Initial value of the effective temperature $\chi_0$ as a function of bath temperature]{Initial value of the effective temperature $\chi_0$ as a function of bath temperature for a material obeying the STZ equations with effective temperature with the same parameters as Fig.~\ref{fig:s_vs_strain:most_unstable} after being cooled from $T_{initial} = T_{eff} = T_{inf}$ to 295 K at 1 K/s, then, some time later, heated to the final temperature $T$ at 20 K/min, and held at that temperature for 10 minutes.}
  \label{fig:chi_initial}
\end{figure}

Immediately before testing, the samples are heated to the testing temperature and annealed for a few minutes.  During the annealing process, if the final temperature is high enough, the effective temperature will have time to equilibrate to the bath temperature.  Figure~\ref{fig:chi_initial} shows the final effective temperature as a function of the bath temperature for a system following a thermal history similar to that performed in \cite{JLGR03}, cooled from the melt to room temperature, held there for more than a few hours,
 then heated at a rate of 20 K/min to a final temperature at which they are anneled for 10 minutes.
At high temperatures, the system completely equilibrates, and the thermal history is forgotten;
at low temperatures, the thermal effects are still not strong enough to noticeably change the temperature; and in between the material is equilibrating over laboratory timescales.

\begin{figure}
  \centering
  \includegraphics[width=\columnwidth,clip]{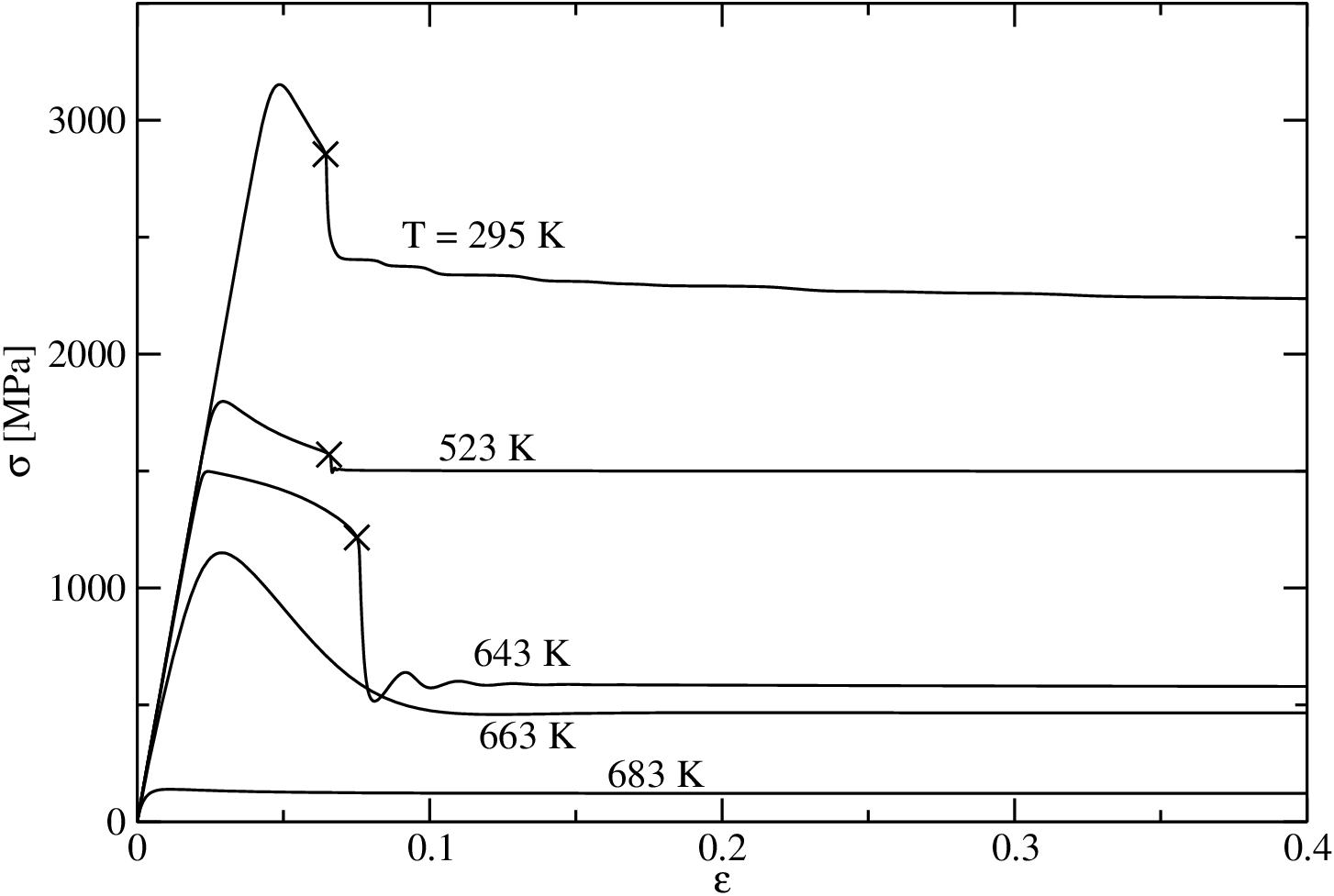}
  \caption[Stress vs.\ strain for a system initially at rest for varying bath temperatures]{Stress vs.\ strain for a system initially at rest for five different temperatures at a strain rate of $\dot{\varepsilon} = 0.1$ $\mathrm{s}^{-1}$.  All other parameters are the same as in Fig.~\ref{fig:stress_vs_strain}.}
  \label{fig:stress_vs_strain:T_varying}
\end{figure}

Figure~\ref{fig:stress_vs_strain:T_varying} plots the stress as a function of strain for a strain rate of $0.1$ $\mathrm{s}^{-1}$ at five different temperatures with the same parameters as used in Fig.~\ref{fig:stress_vs_strain}, except for the initial effective temperature, which is taken from Fig.~\ref{fig:chi_initial}.  Note that the initial heterogeneities are calculated as a percentage of the initial effective temperature; if a point is $0.25\%$ above the average at $T=295$ K, it is also $0.25\%$ above the average for $T=683$ K.

Qualitatively, the model agrees well with experiment.  Localization is only seen in at the lowest three temperatures, occurring when the stress sharply drops.  Quantitatively, the peaks are higher than seen in the experiments\cite{JLGR03}, $3150$ MPa from the model versus $1850$ MPa in the experiments, but this is the regime of low temperatures and high strain rates where the rate factors are most likely to be incorrect.  This though is much better agreement than if the effective temperature was taken to be the actual bath temperature.  In that case, the initial number of zones for the lowest two temperatures, the stresses become outrageously bigger than in the experiments.  Roughly $5{,}000$ MPa for $T = 523$ K and $25{,}000$ MPa for $T = 295$ K.  Trying to eliminate the effective temperature by simply using the bath temperature will not work.

\begin{figure}
  \centering
  \includegraphics[width=\columnwidth,clip]{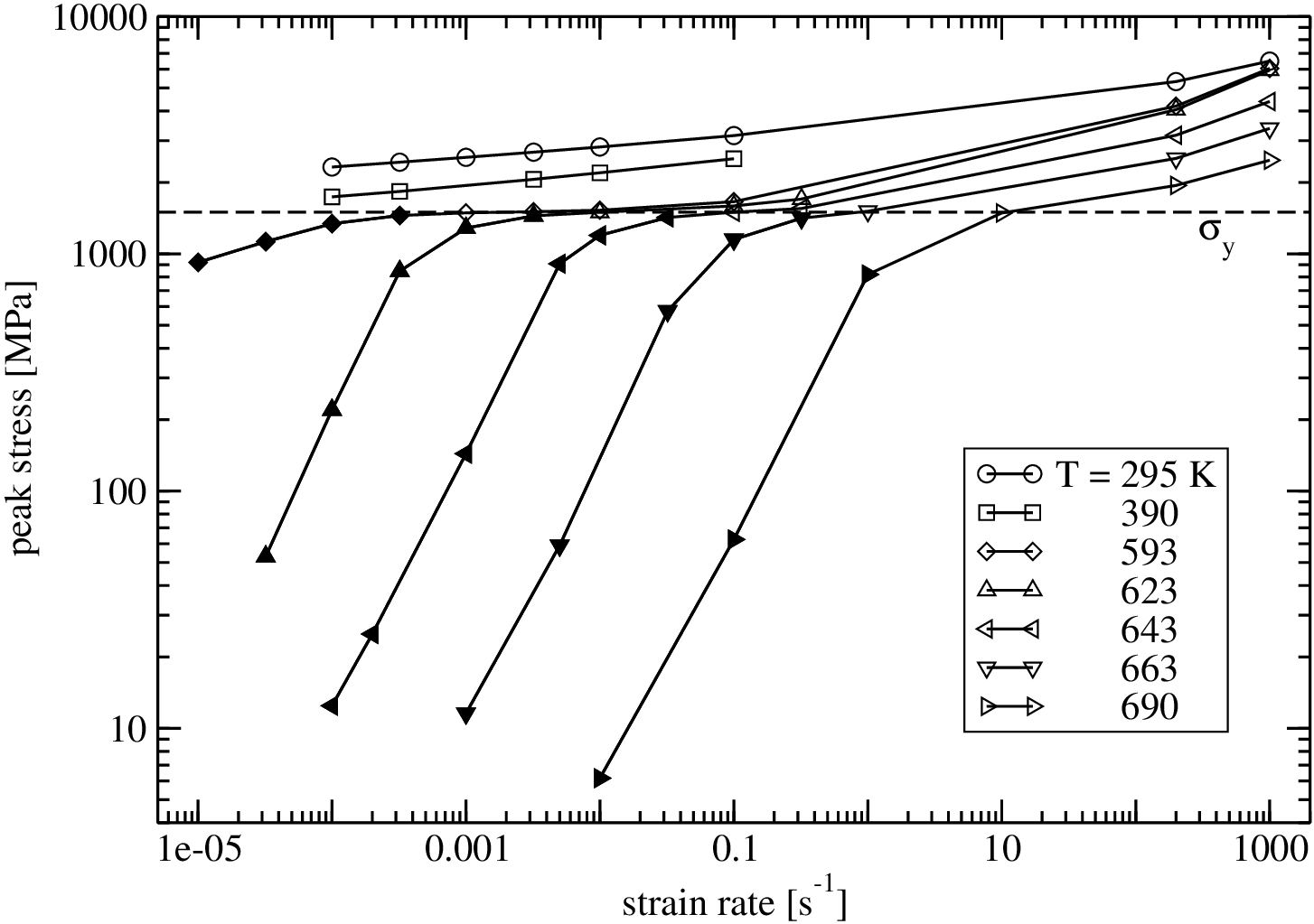}
  \caption[Peak stress vs.\ applied strain rate at various temperatures]{Peak stress vs.\ applied strain rate at various temperatures for STZ models initially at rest.  The dashed line is the STZ yield stress.}
  \label{fig:peak_stress_vs_strain_rate}
\end{figure}

Figure~\ref{fig:peak_stress_vs_strain_rate} shows the peak stress vs.\ strain rate for a number of different temperatures.  Solid symbols are those that remain localized up to the final strain; open symbols exhibit strain localization.  The logic used earlier predicts that those runs where the peak stress goes above 1 should localize and the numerical results support this.  All the runs with peak stress below 0.974 stay uniform; all the runs with peak stress above 0.99 localize.  (There were no results in between.)  Any samples whose peak stress rises above this value localize, while any materials where the stress stays below it remain uniform.  When the strain rate is low and the temperature is high, the response is dominated by linear viscosity, as expected.
For the higher stresses, the material is forced at a rate higher than the creep alone can deliver.  Instead, the flow is dominated by the STZ flow, and the temperature becomes irrelevant.  More important is the initial number of zones.

\begin{figure}
  \centering
  \includegraphics[width=\columnwidth,clip]{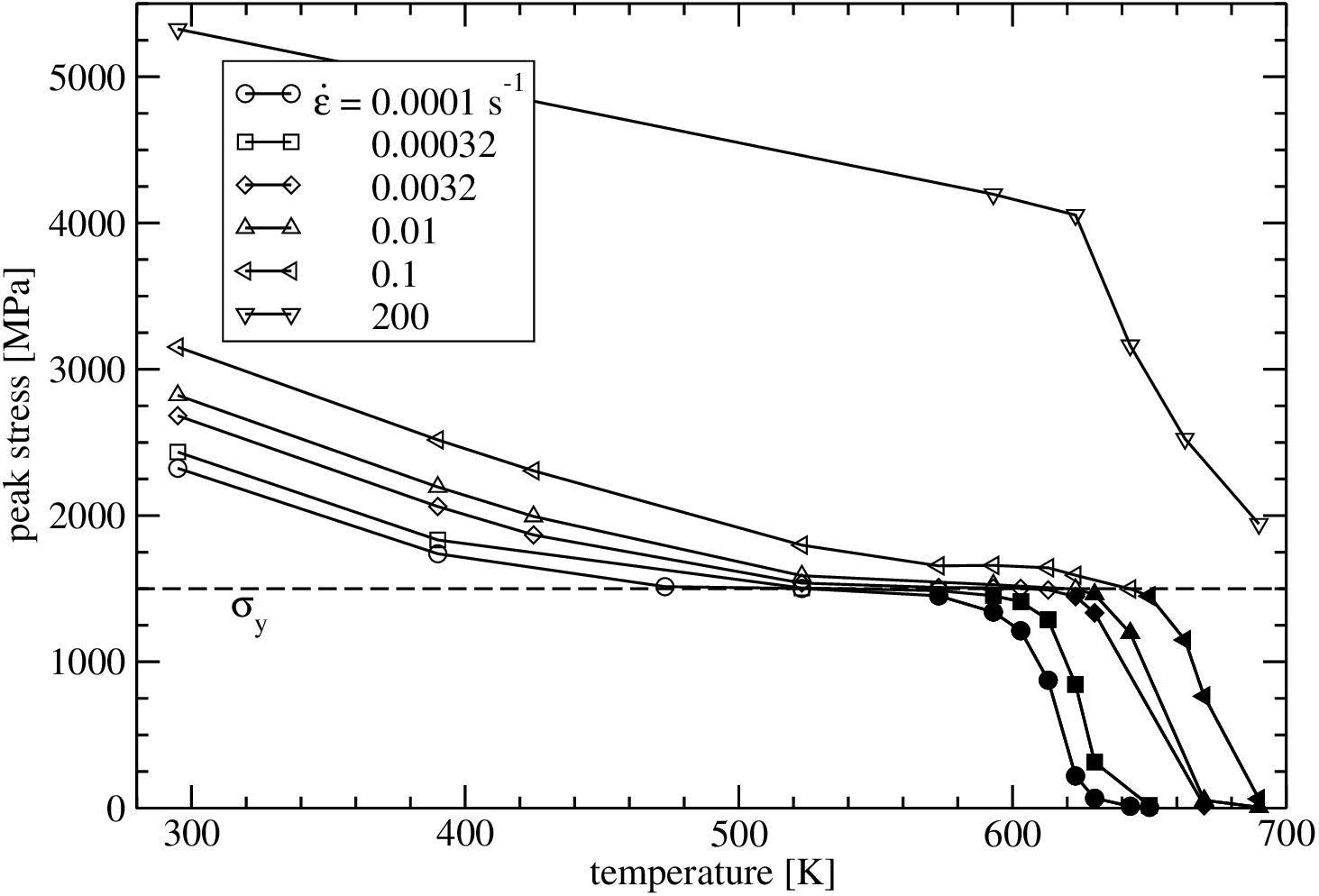}
  \caption[Peak stress vs.\ temperature at various applied strain rates]{Peak stress vs.\ temperature at various applied strain rates for STZ models initially at rest.  The dashed line is the STZ yield stress.}
  \label{fig:peak_stress_vs_temperature}
\end{figure}

Figure~\ref{fig:peak_stress_vs_temperature} shows the peak stress as a function of temperature for varying strain rates.  Again, the change in behavior as the peak stresses rises above the yield stress is seen.  Note that although the data for $\dot{\varepsilon}=200$ $s^{-1}$ in Fig.~\ref{fig:peak_stress_vs_temperature} appears to become viscous at high temperatures, the change in slope is a result of the change in $\chi_0$.  Remember, above 620 K, the initial effective temperature is equal to the bath temperature, so when the temperature increases, not only does $\rho$ increase, but $\chi_0$ does as well.  More zones are available to contribute to the plastic flow, meaning the same strain rate can be attained at a lower stress.

\begin{figure}
  \centering
  \includegraphics[width=\columnwidth,clip]{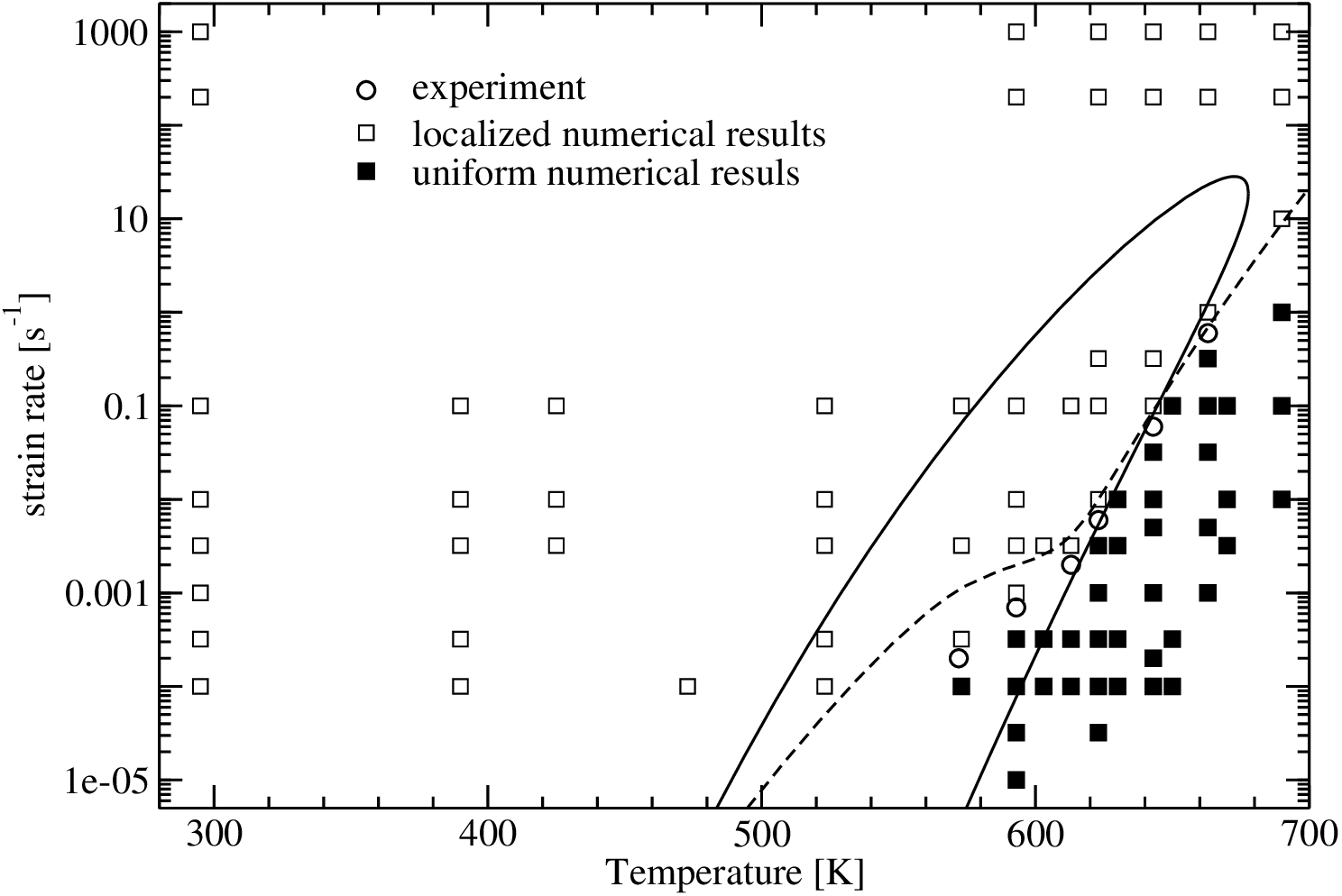}
  \caption[Scatter plot of temperature and applied strain rate pairs for the STZ model started at rest]{Scatter plot of temperature and applied strain rate pairs for the STZ model starting at rest ($s_0 = 0$, $m_0 = 0$ and $\chi_0$ according to Fig.~\ref{fig:chi_initial}).  The open squares correspond to runs that exhibit localization while the solid squares remained uniform.  The circles are experimental data.  The solid line is the boundary of instability under infinitely small perturbation.  The dashed line is $\dot{\varepsilon}_{max creep} (T)$.  Below it, the flow is primarily creep, and above the flow is primarily from the STZ dynamics.}
  \label{fig:localization_scatter_plot_for_temperature_and_strain_rate_runs}
\end{figure}

Figure~\ref{fig:localization_scatter_plot_for_temperature_and_strain_rate_runs} is a scatter plot of the peak stress at many different pairs of temperature and applied strain rate.  The circles are the experimental strain rates below which the system is uniform and above which the system localizes taken from \cite{JLGR03}.  Inside the solid line is the region in which the steady state of the model is non-uniform.  The solid squares correspond to systems in which the strain rate remains uniform up to a strain of 0.45; the open squares are those systems where the deformation localizes.  The data shows that the existence of a shear band is highly dependent on the initial conditions.  For example, while all the points tested below 500 K should be uniform in the steady state
, they all localize as the system responds to the applied strain rate.  Instead of a a uniform increase in effective temperature towards the final uniform solution, the material creates one large shear band first where the effective temperature is greatest.  That shear band allows the stress to drop, slowing the heating in the remainder of the system.  Instead, the system is heated by diffusion of effective temperature from the shear band and the band widens to absorb the entire sample.  The numerical equations appeared too stiff to verify the eventual return to homogeneous flow at that temperature, but it was verified for $T=690$~K and $\dot{\varepsilon} = 200$~$\mathrm{sec}^{-1}$ where $\phi$ drops below 0.5 at a strain of roughly 15.  Likewise, the point $T = 593$~K and $\dot{\varepsilon} = 3.2 \times 10^{-4}$~$\mathrm{sec}^{-1}$ which was initially uniform by my criteria, slowly but steadily localizes, reaching $\phi = 0.5$ at a strain of 0.9.

If the logic proposed earlier holds true, those strain rates that are low enough that they can be produced solely by the thermal creep from just the initial density of zones should not show localization in experiment.  In other words, for $\dot{\varepsilon}$ such that
\begin{equation}
  \label{eq:33}
  \dot{\varepsilon} < \dot{\varepsilon}_{max creep}(T) = \frac{\varepsilon_0}{\tau_0} e^{-1/\chi_0(T)} \mathcal{C}(1) \left( \mathcal{T}(1) - m_{ss}(1) \right),
\end{equation}
the material should appear to remain uniform, while materials forced at strain rates $\dot{\varepsilon} > \dot{\varepsilon}_{max creep}(T)$ should localize.  $\dot{\varepsilon}_{max creep}(T)$ is the dotted line and it agrees well with the numerical results, only differing for those points where the max stress is between 0.974 and 0.99 $s_y$.

\section{Conclusion}
\label{sec:conclusion}

  As shown in the previous sections, the addition of the effective temperature dynamics into the STZ theory enables it to qualitatively describe the localization phenomenology seen in experimental data.

The shear banding caused by this characteristic of the model is different from that proposed in \cite{JSL01}.  In that paper, the bands were the result of the interactions between STZ flips mediated by the elasticity of the material.  The resulting shear bands were characterized by differences in the STZ bias, $\Delta = m \Lambda$.  In contrast, these shear bands are caused solely by inhomogeneities in the density, not bias of STZs.  The previously proposed mechanism leads to bands of a definite wavelength, while the mechanism discussed here does not.  The wavelength undetermined, but the number of bands itself can vary depending on the initial conditions.  

In addition to an instability of the steady state, the transient response to initial loading can develop shear bands.  When a system where the effective temperature is noisy is loaded at a constant strain rate, the ``hotter'' regions, because they have more zones, experience a disproportionate share of the plastic strain.  More deformation means more dissipation from plastic work, which means the effective temperature increases faster, which means more zones, which allows for greater plastic deformation.  As a result, the deformation can possibly localize in some regions, while the remainder of the material is relatively undeformed.

This instability in the response of the material to the initial loading was shown to well describe the strain rates and temperatures for which localization appears in the experiments.  Since this dynamical instability map does not necessarily require instabilities of uniform steady-state results, such an instability might exist in the original $\beta = 1$ version of the model Langer studied previously.  That version does fit the steady-state stress vs.\ strain rate data better and is a simpler form in which to compute results analytically.  Such a study is an obvious avenue for future research.

\section{Acknowledgments}
\label{sec:acknoledgments}

I would like to thank Craig Maloney and Lisa Manning for useful discussions and especially Jim Langer for his guidance and encouragment.  This research was supported by U.S.\ Department of Energy Grant No.\ DE-FG03-99ER45762 and by the MRSEC Program of the National Science Foundation under Award No.\ DMR0080034.

\bibliography{JabRef}

\end{document}